\documentclass[12pt,a4paper]{article}

   \usepackage{a4wide}
    \usepackage{subcaption}
    \usepackage[english]{babel}
    \usepackage{amsmath}   
    \usepackage{graphicx}  
    \usepackage{dcolumn}   
    \usepackage{bm}        
    \usepackage{amssymb}   
    \usepackage{epsfig}
    \usepackage{epstopdf}
    \usepackage{mdframed}
    \usepackage{amsmath}
    \usepackage{wasysym}
    \usepackage[toc]{appendix}
    \usepackage{color,soul} 
    \usepackage{datetime}
    \usepackage{physics}
    \usepackage{enumerate}
    \usepackage{tablefootnote}
    \usepackage{multirow}
    \usepackage{empheq}
    \usepackage{hhline}
    \usepackage{esint}
    \usepackage{tensor}
    
    \usepackage{footmisc}
    
    \usepackage{hyperref}

    \newcommand{\be}{\begin{equation}}
    \newcommand{\ee}{\end{equation}}
    \newcommand{\bea}{\begin{eqnarray}}
    \newcommand{\eea}{\end{eqnarray}}
    \newcommand{\bean}{\begin{eqnarray*}}
    \newcommand{\eean}{\end{eqnarray*}}
    \newcommand{\nn}{\nonumber}

    \newcommand{\ttheta}{\tilde{\theta}}

    \def \A {\mathcal{A}}
\def \B {\mathcal{B}}
\def \C {\mathcal{C}}
\def \D {\mathcal{D}}
\def \E {\mathcal{E}}
\def \F {\mathcal{F}}
\def \G {\mathcal{G}}
\def \T {\mathcal{T}}
\def \F {\mathcal{F}}
\def \G {\mathcal{G}}
\def \H {\mathcal{H}}
\def \I {\mathcal{I}}
\def \J {\mathcal{J}}
\def \g {\mathfrak{g}}

\def \S {\mathcal{S}}

\def \O {\mathcal{O}}
\def \l {\ell}

\def \Jone {J^{(1)}}
\def \Jtwo {J^{(2)}}

\newcommand*\Laplace{\mathop{}\!\mathbin\bigtriangleup}

\begin{document}

\begin{titlepage}

\numberwithin{equation}{section}
\begin{flushright}
\small

\normalsize
\end{flushright}

\begin{center}

{\LARGE \textbf{Gravitational Multipoles in Five Dimensions}}
\\

\medskip

\vspace{1 cm} {\large Jef Heynen$^{1}$, Daniel R. Mayerson$^1$}\\

\vspace{1cm}
 
 {$^1$ Institute for Theoretical Physics, KU Leuven, Celestijnenlaan 200D, B-3001 Leuven, Belgium}

\vspace{1cm}

jh2467 @ cam.ac.uk , daniel.mayerson @ kuleuven.be

\vspace{1cm}

\textbf{Abstract}
\end{center}

\noindent We define gravitational mass and current multipoles for five-dimensional, stationary, and asymptotically flat vacuum metrics. We do this by generalizing Thorne's asymptotically Cartesian and mass-centered (ACMC) coordinate formalism to five dimensions, and prove that the multipoles defined in this way are unambiguously well-defined.
Further, these two towers of multipole tensors, in the case of biaxial symmetry, reduce to a tower of
 mass multipoles $M_\l$, and two separate towers of current or angular momentum multipoles $S^{(1)}_\l, S^{(2)}_\l$.
We apply our formalism to a few examples, in particular Myers-Perry black holes, black rings, and smooth multicentered geometries.

\vspace{2em}


\setcounter{tocdepth}{2}
\tableofcontents

\end{titlepage}
\newpage
\section{Introduction and Summary}

Gravitational multipoles give a unique way of parametrizing and quantifying the asymptotic gravitational field produced by a massive object; successive multipole orders give more and more precise angular resolution of the object's structure. As such, their presence is ubiquitous in gravitational wave physics \cite{LIGOScientific:2021sio,Cardoso:2019rvt,Ryan:1995wh,Barack:2006pq,Vigeland:2010xe,Mayerson:2020tpn,Fransen:2022jtw,Loutrel:2022ant}, where the multipole structure of observed objects in binary systems can possibly represent a smoking gun of physics beyond general relativity.

Clearly, the most interesting spacetime arena to understand gravitational multipole phenomenology is four dimensions. However, in this paper, we develop the theory and definitions of multipoles in \emph{five} dimensions. This allows us to use multipoles to understand phenomena which may not have an analogue in four dimensions --- perhaps the most obvious example is the no-hair theorems, which only exist in four dimensions \cite{Emparan:2008eg}. In five (or more) dimensions, there can exist multiple black objects with the same mass and angular momenta, violating black hole uniqueness implied by the no-hair theorems of four dimensions. As we show here, higher-order multipoles \emph{do} distinguish between such different black objects, as one would expect. 

\bigskip

Defining gravitational multipoles is a subtle endeavor in gravity due to general coordinate invariance.
Geroch first devised a coordinate-invariant formulation of multipole moments in flat spacetime, where multipole moments were related to conformal tensors  evaluated at spatial infinity \cite{Geroch:1970:1}. He then generalized this formalism to curved, static vacuum spacetimes \cite{Geroch:1970:2}, and Hansen \cite{Hansen:1974} further generalized the formalism to generic stationary (still vacuum) spacetimes as well.\footnote{Note that the three formalisms of Geroch-Hansen, Thorne, and Noether charge multipole formalisms are originally defined for vacuum spacetimes. In certain cases, it was argued that the formalisms could be generalized beyond vacuum --- e.g. for Geroch-Hansen and Thorne, to Einstein-Maxwell theory \cite{Simon:1983,Hoenselaert:1990,Sotiriou:2004}, or more recently for generic non-vacuum spacetimes \cite{Mayerson:2022:oct}; all three formalisms can also be consistently extended to higher-derivative theories \cite{Cano:2022}.}

A few years later, Thorne \cite{Thorne:1980} developed a formalism defining a family of asymptotically Cartesian and mass-centered (ACMC) coordinates, in which multipole moments could unambiguously be read off from the metric coefficients. For stationary spacetimes, G\"ursel \cite{Gursel:1983} showed that the Geroch-Hansen and Thorne multipoles were equal (up to an unimportant normalization), but Thorne's formalism had the advantage of also being applicable to non-stationary, radiating spacetimes, with time-dependent multipole moments.

More recently, an elegant formalism (also equivalent to Thorne's) was developed that defined multipole moments as Noether charges associated with asymptotic multipole symmetries \cite{Compere:2018}; this formalism has the advantage of being more readily generalizable to certain scenarios such as asymptotically de Sitter spacetimes \cite{Chakraborty:2021ezq}.

\bigskip

In five dimensions, there has been very little attempt to seriously and rigorously define multipoles in any way, although their existence (and form) have sometimes been assumed in the study of hypothetical five-dimensional gravitational observables, see e.g. \cite{Cardoso:2019vof}. To the best of our knowledge, the only attempt to carefully define multipole moments in five dimensions was a higher-dimensional generalization  of the Geroch-Hansen formalism by Tanabe, Ohashi, and Shiromizu \cite{Tanabe:2010}. This definition was incomplete in various ways (although the specific apparent ambiguity they reported in defining the Geroch-Hansen multipoles beyond quadrupole order is not actually present); we discuss this further in detail in Appendix \ref{app:GH}.

In our paper, we  define mass and current gravitational multipole moments for stationary, asymptotically flat vacuum solutions to Einstein's equations, using a five-dimensional generalization of Thorne's ACMC formalism. As we show, this formalism is readily generalizable to five dimensions and allows the unambiguous definition of multipole moments (as we show) for such spacetimes.

\bigskip

The rest of this paper is structured as follows.
In the following Section \ref{sec:summary}, we briefly summarize the main results of our work. In Section \ref{sec:linear}, we solve the five-dimensional vacuum Einstein equations in the de Donder gauge, first in linearized gravity and then arguing what the form of the non-linear completion must be. 
These results lead naturally to our definition of ACMC coordinate systems in Section \ref{sec:ACMC}, for which we give the explicit definitions in terms of asymptotic expansions of the metric coefficients, and also give the explicit expressions in the case where the metric additionally enjoys biaxial symmetry.
In Section \ref{sec:examples}, we apply this new five-dimensional ACMC formalism to a few examples (the Myers-Perry black hole, the black ring with both one and two angular momenta, and supersymmetric smooth multicentered geometries).
Finally, Appendix \ref{app:STF} gives an overview of STF tensors and also discusses the new ASTF tensors that we have introduced, and Appendix \ref{app:GH} discusses the (only) earlier attempt to define multipoles in five dimensions using the Geroch-Hansen formalism.

\bigskip

\textbf{Note added in v2:} The authors of \cite{Gambino:2024uge} pointed out that, in addition to the towers of mass and current multipoles that we discuss, a third  type of multipole tower exists in five spacetime dimensions and was incorrectly missed in our analysis. This third multipole tower is associated to the purely spatial components of the metric and are coined ``stress multipoles'' in \cite{Gambino:2024uge}.  In this version of our paper, we indicate the places in which these stress multipoles would feature and alter the equations listed. Note that our discussion of mass and current multipoles, and their well-definedness, is not altered (as we indicate explicitly). For the definition of stress multipoles (and a calculation of their value for certain metrics), see \cite{Gambino:2024uge}.

\subsection{Summary}\label{sec:summary}
We define a family of asymptotically Cartesian and mass-centered (ACMC) coordinates for stationary, vacuum spacetimes in five dimensions. These are discussed and defined analogously to Thorne's original four-dimensional treatment of ACMC coordinates  \cite{Thorne:1980}, first constructing solutions to the linearized Einstein equations,\footnote{Note that our linearized solutions do not include the ``stress multipole'' terms defined and discussed in \cite{Gambino:2024uge}.} arguing for the general form of its non-linear completion, and finally defining a more general family of ACMC coordinates in which the multipole moments can be read off from the coefficients of the asymptotic radial expansion at infinity.

\bigskip

In particular, our main results are  expressions for such a stationary, vacuum five-dimensional spacetime in ACMC coordinates given in \eqref{eq: CH3: definition ACMC coordinates 5D}. We also give the reduced expressions for the most prevalent situation of biaxial spacetimes (which also admit two commuting spacelike Killing vectors with closed orbits) in \eqref{eq: CH3: ACMC conditions 5D axisymmetry}.
These expansions define the tower of mass multipole moment tensors $\g_{A_\l}$ (or simply $M_\l$ in the case of biaxial symmetry) and the tower of current multipole tensors $\S_{bc\, A_{\l}}$ (which reduce to two towers of coefficients $S^{(1)}_\l, S^{(2)}_\l$ for biaxial symmetry).
Crucially, we show (in Section \ref{sec:proofinvariance}) that these definitions of multipole moments are indeed unambiguously defined, in that they do not depend on the particular ACMC coordinate system used.

\bigskip
We exhibit how the ACMC formalism can be used in practice to calculate mass and current multipole moments of spacetimes in Section \ref{sec:examples}. The simplest non-trivial example is the Myers-Perry black hole, resulting in the (biaxially symmetric) non-zero multipole moments (see \eqref{eq:MPBHmultipoles}):
\begin{equation*}
    M_{2n}=(a^2-b^2)^{n}M\,,\quad S_{2n+1}^{(1)}=\frac{2n+1}{n+1}(a^2-b^2)^nJ^{(1)}\,,\quad S_{2n+1}^{(2)}=\frac{2n+1}{n+1}(a^2-b^2)^nJ^{(2)}\,,
\end{equation*}
where $J^{(1)} = (2/3)Ma$ and $J^{(2)} = (2/3)Mb$.
This can be compared to four-dimensional Kerr (with parameters $M,a$) which has non-zero multipoles $M_{2n} = M(-J^2/M^2)^n$ and $S_{2n+1} = J(-J^2/M^2)^n$.

We also calculate mass and current multipole moments for the vacuum black ring with one or two angular momenta; see Section \ref{sec:BRmultipoles}. Such black rings can carry the same mass and angular momenta as Myers-Perry black holes; we illustrate that the higher multipole moments can be used as asymptotic observables that break the degeneracy of vacuum solutions with the same mass and angular momenta. (This was also a key point illustrated in \cite{Tanabe:2010} using the Geroch-Hansen formalism. We discuss their attempt at generalizing multipole moments to five dimensions in some detail in Appendix \ref{app:GH}.)

Finally, we also illustrate that the ACMC formalism developed here --- even though in principle only defined for vacuum spacetimes --- can also be used for non-vacuum spacetimes (analogously to the four-dimensional case, as used in e.g. \cite{Bena:2020see,Bianchi:2020bxa,Bena:2020uup,Bianchi:2020miz,Bah:2021jno}). In particular, we calculate the multipole moments of five-dimensional supersymmetric smooth multicentered Bena-Warner geometries in Section \ref{sec:BWmultipoles}. (Note that the four-dimensional multipoles of these geometries were calculated in \cite{Bena:2020uup} for choices of moduli that were appropriate for four-dimensional asymptotics.)

\bigskip

There are numerous possible future directions in the program of five-dimensional multipole moments. The most obvious such directions are to extend all of the different existing four-dimensional multipole formalisms to five dimensions.
First of all, it would certainly be interesting to investigate how the Geroch-Hansen formalism generalizes to five dimensions --- as we discuss in  in Appendix \ref{app:GH}, the work of \cite{Tanabe:2010} is certainly a large step in the right direction, but more steps are required to make the multipoles in this formalism rigorously well-defined. Moreover, one could then presumably show that the equivalence between the Geroch-Hansen and Thorne formalisms continues to hold. 
The ACMC formalism itself can be extended to include time-dependent spacetimes and multipole moments (as Thorne's original work does in four dimensions \cite{Thorne:1980}). Finally, the Noether charge formalism of \cite{Compere:2018} could also presumably be generalized to five dimensions, and should also naturally be applicable to spacetimes beyond stationarity.

\section{Multipoles in de Donder Gauge}\label{sec:linear}

In five spacetime dimensions, we will use $x^0=t$ to denote the time direction and $x_i$ with $i=1,2,3,4$, as the standard four Cartesian coordinates. We will also use the spherical (Hopf) coordinates:
\begin{align}
    \label{eq: CH3: definition Hopf coordinates}
    &x_1=r\sin\theta\cos\phi_1\,,\,
    x_2=r\sin\theta\sin\phi_1\,,\,
    x_3=r\cos\theta\cos\phi_2\,,\,
    x_4=r\cos\theta\sin\phi_2\,,\\\nonumber
    &\theta\in [0,\tfrac{\pi}{2}]\,,\quad \phi_1,\phi_2\in[0,2\pi)\,.
\end{align}
The flat spacetime metric is then given by:
\be\label{eq:flatspace} ds^2 = -dt^2 + \sum_{i=1}^4 dx_i^2 = -dt^2 + dr^2 + r^2d\theta^2 + r^2\sin^2\theta\, d\phi_1^2 + r^2\cos^2\theta\, d\phi_2^2.\ee
We will denote $n_j = x_j/r$ as the radial unit vector, and will often abbreviate a sequence of indices as
\be \A_{A_\l} \equiv \A_{a_1 a_2\cdots a_{\l}}.\ee
A product of $\l$ radial unit vectors will also be abbreviated as:
\be N_{A_\l} \equiv n_{a_1}\cdots n_{a_\l}.\ee
Note that $N_{A_\l}$ represents an angular dependence of order $\l$; when expressed in terms of the spherical coordinates $(r,\theta,\phi_i)$, it is independent of $r$ and can be written as a polynomial of (at most) degree $\l$ in $\sin\theta,\cos\theta,\sin\phi_i,\cos\phi_i$.
 
Note that the flat Laplacian $\Laplace$ in four spatial dimensions admits the solution:
\be \Laplace\left(\frac{1}{r^2}\right) = 0,\ee
as opposed to the $1/r$ solution in three spatial dimensions. It follows that the most general (scalar) solution to $\Laplace f = 0$ that vanishes at infinity can be written as:
\be f = \sum_{\l=0}^\infty \F_{A_\l} \partial_{A_\l} \left( \frac{1}{r^2}\right) = \sum_{\ell=0}^\infty \F_{A_\l} \left( \frac{1}{r^2}\right)_{,\,A_\l} ,\ee
where, by construction, $\F_{A_\l}$ are constant, symmetric and trace-free (STF) tensors, i.e. $\F_{a_1 a_2\cdots a_\l}$ = $\F_{(a_1 a_2\cdots a_\l)}$ and $\F_{a_1 a_1\cdots a_\l}=0$.

In our discussion, we will also need a new type of tensor $\T_{bc A_\l} = \T_{bc\, a_1\cdots a_\l}$,  which is STF on all indices except the first two, and is antisymmetric on the first two indices, $\T_{bc\, a_1\cdots a_\l} = \T_{[bc]\, a_1\cdots a_\l}$. Additionally, all traces vanish, $\T_{ac\, a a_2\cdots a_\l} = 0$, and any antisymmetrization over three indices vanishes, i.e. for all $i$, $\T_{[bc\, |a_1\cdots a_{i-1}|a_i] a_{i+1} \cdots a_\l} = 0$. We call this type of tensor ``anti-symmetric + STF'' (ASTF) and discuss them in more detail in Appendix \ref{app:STF}. ASTF tensors arise naturally in the discussion of angular momentum or current multipoles, as we will see below; the reason these do not appear in Thorne's original four-dimensional analysis is that a pair of antisymmetric indices can be exchanged for a single index in three spatial dimensions.

\subsection{Linearized theory}
We expand the spacetime metric linearly around flat space as:
\begin{equation*}
    g_{\alpha\beta}=\eta_{\alpha\beta}+g^1_{\alpha\beta}\,.
\end{equation*}
It is more convenient to work with the trace-reversed perturbation $\gamma^1_{\alpha\beta}$:
\begin{equation}
    \label{eq: CH3: definition trace-reversed perturbation}
    \gamma^1_{\alpha\beta}=g^1_{\alpha\beta}-\frac{1}{2}\eta_{\alpha\beta}\eta^{\mu\nu}g^1_{\mu\nu}\,.
\end{equation}
Note that the inverse of this relation is:
\begin{equation}
    \label{eq: CH3: inverse relation perturbation and trace-reversed}
    g^1_{\alpha\beta}=\gamma^1_{\alpha\beta}-\frac{1}{3}\eta_{\alpha\beta}\eta^{\mu\nu}\gamma^1_{\mu\nu}\,.
\end{equation}
which is slightly different than its four-dimensional counterpart. After choosing the Lorentz gauge
\begin{equation*} \label{eq:lorentzgaugelinear}
    {\gamma^1_{\alpha\beta\,,\,}}^{\beta}=0\,,
\end{equation*}
and demanding stationarity, $\partial_t \gamma^1_{\alpha\beta} = 0$, the vacuum Einstein equations reduce to the simple Laplacian equations:
\begin{equation}
  \label{eq:vacuumlinear}  \gamma_{\alpha\beta\,,\,jj}^{1}=\Laplace \gamma_{\alpha\beta}^1=0\,,
\end{equation}
The most general solution (without stress multipoles\footnote{Our analysis erroneously neglects the ``Riemann-symmetric term'' which leads to a new, third tower of ``stress multipoles'' as discussed in \cite{Gambino:2024uge}. Our discussion can be thought of as focusing only on the mass and current multipoles.}) to (\ref{eq:vacuumlinear}) is given by:
\begin{align}
    \gamma^1_{00}&=\sum_{\l=0}^\infty \left[\A_{A_\l}-\E_{A_\l}\right]\left(\frac{1}{r^2}\right)_{,\,A_\l}\\[1em]
    \label{eq:mostgenerallaplacesolvec}
    \gamma^1_{0j}&=\sum_{\l=0}^\infty\B_{A_L}\left(\frac{1}{r^2}\right)_{,\,jA_\l}+\sum_{\l=1}^\infty\C_{jA_{\l-1}}\left(\frac{1}{r^2}\right)_{,\,A_{\l-1}}+\sum_{\l=2}^\infty \varepsilon_{jbca}\D_{bcA_{\l-2}}\left(\frac{1}{r^2}\right)_{,\,a A_{\l-2}}\\[1em] \nonumber
    \label{eq:mostgenerallaplacesoltensor}
    \gamma^1_{ij}&=\sum_{\l=0}^\infty\left(\delta_{ij}\E_{A_\l}\left(\frac{1}{r^2}\right)_{,\,A_\l}+\J_{A_\l}\left(\frac{1}{r^2}\right)_{,\,ijA_\l}\right)+\sum_{\l=1}^\infty \G_{(j|A_{\l-1}}\left(\frac{1}{r^2}\right)_{,\,|i)A_{\l-1}}\\\nonumber
    &\quad+\sum_{\l=2}^\infty\left(\F_{(ij)A_{\l-2}}\left(\frac{1}{r^2}\right)_{,\,A_{\l-2}}+\varepsilon_{ibca}\I_{bcA_{\l-2}}\left(\frac{1}{r^2}\right)_{,\,ja A_{\l-2}}\right)\\
    &\quad+\sum_{\l=3}^\infty\varepsilon_{a bc(i}\H_{|bc|j) A_{\l-3}}\left(\frac{1}{r^2}\right)_{,\,a A_{\l-3}}
\end{align}
where $\A_{A_\l}, \B_{A_\l}, \C_{A_\l}, \E_{A_\l}, \J_{A_\l}, \G_{A_\l}, \F_{A_\l}$ are STF tensors and $\D_{bc A_\l}, \H_{bc A_\l}, \I_{bc A_\l}$ are ASTF tensors introduced above (and discussed further in Appendix \ref{app:STF}).
The Lorentz gauge conditions:
\begin{equation}
    \label{eq: CH3: gauge conditions trace-reversed perturbation}
    \gamma^1_{0j\,,\,j}=0\,,\qquad\gamma^1_{ij\,,\,j}=0\,,
\end{equation}
further imply that 
\begin{equation*}
    \C_{A_\l}=0\,, \quad \E_{A_\l}=-\frac{1}{2}\G_{A_\l}\,,\quad \F_{A_\l}=0\,,\quad \H_{A_\l}=0\,,
\end{equation*}
so that the most general solution (without stress multipoles \cite{Gambino:2024uge}) to (\ref{eq:vacuumlinear}) in the Lorentz gauge (\ref{eq:lorentzgaugelinear}) is:
\begin{align*}
\gamma^1_{00}&=\sum_{\l=0}^\infty \left[\A_{A_\l}-\E_{A_\l}\right]\left(\frac{1}{r^2}\right)_{,\,A_\l}\\[1em]
    \gamma^1_{0j}&=\sum_{\l=0}^\infty\B_{A_\l}\left(\frac{1}{r^2}\right)_{,\,jA_\l}+\sum_{\l=2}^\infty \varepsilon_{jbca_\l}\D_{bc A_{\l-2}}\left(\frac{1}{r^2}\right)_{,\,a_\l A_{\l-2}}\\[1em]\nonumber
    \gamma^1_{ij}&=\sum_{\l=0}^\infty\delta_{ij}\E_{A_\l}\left(\frac{1}{r^2}\right)_{,\,A_\l}+\sum_{\l=2}^\infty\varepsilon_{(i|bca_\l}\I_{bcA_{\l-2}}\left(\frac{1}{r^2}\right)_{,\,|j)a_\l A_{\l-2}}
    -2\sum_{\l=1}^\infty \E_{(i|A_{\l-1}}\left(\frac{1}{r^2}\right)_{,\,|j)A_{\l-1}}\\&+\sum_{\l=0}^\infty \J_{A_\l}\left(\frac{1}{r^2}\right)_{,\,ijA_\l}\,.
\end{align*}
Under the  gauge transformation $x^\mu\mapsto x^\mu+\xi^\mu$, we have
\begin{equation*}
\gamma^1_{\alpha\beta}\mapsto \gamma^1_{\alpha\beta}+\xi_{\alpha,\,\beta}+\xi_{\beta,\,\alpha}-\eta_{\alpha\beta}\xi\indices{^\mu_{\,,\,}_\mu}\,.
\end{equation*}
so the Lorentz gauge is preserved as long as $\Laplace \xi^\mu = 0$. We can use this gauge freedom to choose
\begin{align*}
    \xi_0&=-\sum_{\l=0}^\infty \B_{A_\l}\left(\frac{1}{r^2}\right)_{,\,A_\l}\\
     \xi_j&=\sum_{\l=1}^\infty \E_{jA_{\l-1}}\left(\frac{1}{r^2}\right)_{,\,A_{\l L-1}}-\frac{1}{2}\sum_{\l=2}^\infty\varepsilon_{jbca_\l}\I_{bcA_{\l-2}}\left(\frac{1}{r^2}\right)_{,\,a_\l A_{\l-2}}-\frac{1}{2}\sum_{\l=0}^\infty \J_{A_\l}\left(\frac{1}{r^2}\right)_{,\,jA_{\l}}\,.
\end{align*}
and finally obtain:
\begin{align*}
    \gamma^1_{00}&=\sum_{\l=0}^\infty \A_{A_\l}\left(\frac{1}{r^2}\right)_{,\,A_\l}\\
    \gamma^1_{0j}&=\sum_{\l=2}^\infty \varepsilon_{jbca_\l}\D_{bcA_{\l-2}}\left(\frac{1}{r^2}\right)_{,\,a_\l A_{\l-2}}\\
    \gamma^1_{ij}&=0\,.
\end{align*}

It is convenient to define
\begin{equation*}
    \A_{A_\l}=C_\l\,\g_{A_\l}\,,\quad \D_{A_\l}=C_\l'\,\S_{A_\l}\,,
\end{equation*}
for some as-yet unspecified $\l$-dependent constants $C_\l$ and $C_\l'$. In terms of these and using \eqref{eq: CH3: inverse relation perturbation and trace-reversed}, the linear metric perturbation we have found is
\begin{align}
    \label{eq: CH3: metric components to linear order g_00}
    g^1_{00}&=\frac{2C_0\,\g}{3 r^2}+\sum_{\l=1}^\infty \frac{2 C_\l}{3}\g_{A_L} (r^{-2})_{,\,A_\l}\\
    \label{eq: CH3: metric components to linear order g_0j}
    g^1_{0j}&=\frac{C_0'\varepsilon_{jbca}\S_{bc}n_a}{r^3}+\sum_{\l=3}^\infty C_\l'\,\varepsilon_{jbca_\l}\S_{bcA_{\l-2}}(r^{-2})_{,\,a_\l A_{\l-2}}\\
    \label{eq: CH3: metric components to linear order g_ij}
    g^1_{ij}&=\delta_{ij}\left(\frac{C_0\,\g}{3 r^2}+\sum_{\l=1}^\infty \frac{C_\l}{3}\g_{A_L} (r^{-2})_{,\,A_\l}\right)\,.
\end{align}
Analogously to the four-dimensional analysis, solutions to the linear vacuum Einstein equations contain a tower of (constant) STF tensors $\g_{A_\l}$ and a tower of (constant) ASTF tensors $\S_{bc A_{\l-1}}$; these two towers represent the mass and current multipoles of the solution.\footnote{Note that, as discussed in \cite{Gambino:2024uge}, our solution \eqref{eq: CH3: metric components to linear order g_00}-\eqref{eq: CH3: metric components to linear order g_ij} is not the most general, as the purely spatial components (\ref{eq: CH3: metric components to linear order g_ij}) can also have an additional contribution from a third tower of ``stress'' multipole moments. (These do not exist in four spacetime dimensions.)}

\subsection{Non-linear completion}
Thorne \cite{Thorne:1980} (also using arguments from \cite{Thorne:1975} and \cite{Landau:1975}) argued how the linear vacuum solutions in four dimensions can be non-linearly completed and lead to a specific ACMC-form of the metric. These arguments continue to hold in an arbitrary number of dimensions. Here, we simply sketch these arguments, generalizing to an arbitrary number of dimensions $d$.

Define the field $\overline{\gamma}_{\alpha\beta}$ as 
\begin{equation}
    \label{eq: CH3: definition hbar nonlinear}
    \overline{\gamma}_{\alpha\beta}=\eta_{\alpha\beta}-(-g)^{1/2}g_{\alpha\beta}\,,\quad\text{with}\quad (-g)=-\det \norm{g_{\alpha\beta}}\,,
\end{equation}
then we can write the exact, nonlinear, stationary vacuum Einstein field equations as
\begin{equation}
    \label{eq: CH3: exact, nonlinear, vacuum EFE}
    \Laplace\overline{\gamma}_{\alpha\beta}=W_{\alpha\beta}\,,
\end{equation}
where $W_{\alpha\beta}$ is a source term that is at least quadratic in $\overline{\gamma}_{\alpha\beta}$ (it is   given explicitly in \cite{Thorne:1975}\footnote{Although \cite{Thorne:1975} and \cite{Landau:1975} are interested in the case of four spacetime dimensions, there is no explicit use of the dimensionality of spacetime in the original computation of \cite{Landau:1975} so that \eqref{eq: CH3: exact, nonlinear, vacuum EFE} still holds in arbitrary spacetime dimensions.}).  
We then use \eqref{eq: CH3: definition hbar nonlinear} to write
\begin{equation*}
    -\det \norm{(-g)^{1/2}g_{\alpha\beta}}=(-g)^{\frac{d-2}{2}}
\end{equation*}
Using the nonlinearity expansions:
\begin{equation*}
\overline{\gamma}_{\alpha\beta}=\sum_{p=1}^\infty G^p\overline{\gamma}_{\alpha\beta}^p\,,\quad g_{\alpha\beta}=\eta_{\alpha\beta}+\sum_{p=1}^\infty G^p g^p_{\alpha\beta}\,,
\end{equation*}
the metric determinant can be expressed as:
\begin{equation*}
    (-g)=-\left(\det \norm{(-g)^{1/2}g_{\alpha\beta}}\right)^{\frac{2}{d-2}}=\left(-\det\norm{\eta_{\alpha\beta}-\overline{\gamma}_{\alpha\beta}}\right)^{\frac{2}{d-2}}=1-\frac{2G}{d-2}\eta^{\alpha\beta}\overline{\gamma}_{\alpha\beta}^1+\O(G^2)\,.
\end{equation*}
Inserting this in \eqref{eq: CH3: definition hbar nonlinear} and equating equal order terms, one finds
\begin{equation*}
    g^1_{\alpha\beta}=\overline{\gamma}_{\alpha\beta}^1-\frac{1}{d-2}\overline{\gamma}_{\mu\nu}^1\eta^{\mu\nu}\eta_{\alpha\beta}\,,
\end{equation*}
so $\gamma^1_{\alpha\beta}$ is indeed the trace-reversed linear perturbation as in \eqref{eq: CH3: definition trace-reversed perturbation}.

Demanding the gauge condition (which preserved the gauge \eqref{eq: CH3: gauge conditions trace-reversed perturbation})
\begin{equation*}
    \overline{\gamma}_{\alpha j\,,\,j}=0
\end{equation*}
and making the nonlinearity expansion
\begin{equation*}
    W_{\alpha\beta}=\sum_{p=2}^\infty G^p w^p_{\alpha\beta}
\end{equation*}
where the expansion of $W_{\alpha\beta}$ begins with $p=2$ because it is at least quadratic in $\overline{\gamma}_{\alpha\beta}$ and where $w_{\alpha\beta}^p$ contains only $\overline{\gamma}^q_{\alpha\beta}$ with $q\leq p-1$. Inserting this in \eqref{eq: CH3: exact, nonlinear, vacuum EFE}, one retrieves 
\begin{align}
    \label{eq: CH2: non-linear part gauge condition}
    {\gamma^p_{\alpha\beta\,,\,}}^\beta&=0&\text{(gauge condition)}\,,\\
    \label{eq: CH2: non-linear part Laplace equation}
    \Laplace \gamma^p_{\alpha\beta}&=w^p_{\alpha\beta}&\text{(Laplace equation)}\,,
\end{align}
where $w^p_{\alpha\beta}$ is a source term depending only on $\gamma^q$ with $q\leq p-1$. Note that for $p=1$, \eqref{eq: CH2: non-linear part gauge condition} is simply the Lorentz gauge \eqref{eq:lorentzgaugelinear} for the linear perturbation; the non-linear generalization \eqref{eq: CH2: non-linear part gauge condition} represents the de Donder gauge:
\begin{equation*}
    \left(g^{\mu\nu}\sqrt{-g}\right)_{,\,\nu}=0\,.
\end{equation*}

The fact that the source term $w^p_{\alpha\beta}$ only depends on $\gamma^q$ with $q\leq p-1$ implies that the nonlinear metric terms $\gamma^p_{\alpha\beta}$ must have a lower angular dependency at the same order of $1/r$ compared to the leading order solution, i.e. the linear perturbation $\gamma^1_{\alpha\beta}$. It immediately follows that, when $g_{\alpha\beta}$ is in the de Donder gauge, the multipole moments $\g_{A_\l}, \S_{bc A_\l}$ will always unambiguously be the \emph{leading} order angular dependence at a specified order in $1/r$ in the asymptotic expansion of $g_{00}$ and $g_{0j}$.

\section{ACMC Systems for Multipoles}\label{sec:ACMC}

In the previous section, we found a stationary solution to linearized Einstein equations in the Lorentz gauge with includes mass and current multipole towers. Moreover, we argued that in its non-linear completion in the de Donder gauge, the multipole moment tensors $\g_{A_\l}$ and $\S_{bc A_{\l-1}}$ can unambiguously be read off from the metric as the coefficients of the leading angular dependence at a given order in this $1/r$ expansion of the metric components $g_{00}$ and $g_{0j}$. This analysis was entirely analogous to the four-dimensional case first discussed by Thorne \cite{Thorne:1980}, and also analogously, we can now define a more general class of coordinate systems in which the metric is not necessarily in the de Donder gauge, but in which the mass and current multipoles can still be read off from the leading order angular dependencies at each order in $1/r$. Such coordinate systems were introduced as asymptotically Cartesian and mass-centered to order $N$ (ACMC-$N$) coordinates by Thorne. We define here a five-dimensional version of these ACMC-$N$ coordinate systems. In an ACMC-$N$ coordinate system, a stationary vacuum solution to Einstein's equations admits the asymptotic expansion:\footnote{Our analysis does not include the ``stress multipoles'' discussed in \cite{Gambino:2024uge}. These stress multipoles appear in the leading order angular dependence in the ACMC expansion of the purely spatial metric components, i.e. they contribute to the $(\l'=\l\text{ poles})$ in $g_{ij}$ (but not to any components in $g_{00}$ or $g_{0j}$). Note that the ACMC form \eqref{eq: CH3: definition ACMC coordinates 5D}  is also sufficient to be able to read off the stress multipoles from the metric (in addition to the mass and current multipoles), as shown in \cite{Gambino:2024uge}.}
\begin{mdframed}[frametitle={ACMC-$N$ coordinates in five dimensions}]
\begin{align}
    \nonumber
    g_{00}&=-1+\frac{8}{3\pi}\frac{\g}{r^2}+\frac{(0\text{ pole})}{r^3}+\sum_{\l=2}^N\frac{8}{3\pi\,r^{\l+2}}\Bigg(\g_{A_\l}N_{A_\l}+\sum_{\l'=0}^{\l-1}(\l'\text{ pole})\Bigg)\\ \nonumber
    &\quad +\frac{8}{3\pi\,r^{N+3}}\Bigg(\g_{A_{N+1}}N_{A_{N+1}}+\sum_{\l'\neq N+1}(\l'\text{ pole})\Bigg)+\O\left(1/r^{N+4}\right)\,,\\\nonumber
    g_{0j}&=\sum_{\l=1}^N\frac{4}{\pi\,r^{\l+2}}\Bigg(\,\varepsilon_{jbca_\l}\S_{bcA_{\l-1}}N_{A_\l}+\Big(\l\text{ pole of the form $\B_{A_{\l-1}}N_{jA_{\l-1}}$ or $\C_{jA_l}\,N_{A_l}$}\Big) \\\nonumber
    &\qquad +\sum_{\l'=0}^{\l-1}(\l'\text{ pole})\Bigg)+\frac{4}{\pi\,r^{N+3}}\Bigg(\varepsilon_{j b c a_{N+1}}\S_{bcA_N}N_{A_{N+1}}+\Big(N+1\text{ pole of the form}\\\nonumber 
    &\qquad\qquad\B_{A_{N}}N_{jA_{N}}\text{or $\C_{jA_{N+1}}N_{A_{N+1}}$}\Big)+\sum_{\l'\neq N+1}(\l'\text{ pole})\Bigg)+\O\left(1/r^{N+4}\right)\,,\\ \label{eq: CH3: definition ACMC coordinates 5D}
    g_{jk}&=\delta_{jk}+\sum_{\l=0}^N\frac{1}{r^{\l+2}}\left(\sum_{\l'=0}^{\l}(\l'\text{ pole})\right)+\frac{1}{r^{N+3}}\left(\text{any angular dependence}\right)+\O\left(1/r^{N+4}\right)\,
\end{align}
\end{mdframed}
A term $(\l'\text{ pole})$ represents any angular dependence of $\l'$-th order, e.g. $\sim N_{A_{\l'}}$. Note that at a given order $1/r^{\ell+2}$ in the expansion of $g_{0j}$, the different order-$\ell$ poles $\S_{bc A_{\l-1}}, \B_{A_{\l-1}}, \C_{jA_\l}$ are always unique and thus can be unambiguously extracted.\footnote{Thorne distinguishes the different $\ell$-poles in $g_{0j}$ in four dimensions by parity, which is not sufficient in five dimensions to distinguish $\S_{bc A_{\l-1}}$ from the other $\l$-poles.} Note that this coordinate system is ``mass-centered'' precisely due to the lack of a $(1\text{ pole})/r^3$ term in $g_{00}$, which corresponds to choosing the origin $r=0$ at the center of mass of the spacetime.

In an ACMC-$N$ coordinate system, all mass and current multipoles up to order $\ell=N+1$ can be read off from the metric expansion of $g_{00}$ and $g_{0j}$ --- specifically, $\g_{A_\l}$ for $\l = 0, \cdots, N+1$ and $\S_{bc A_{\l-1}}$ for $\l=1, \cdots, N+1$. Note that the de Donder gauge discussed above is then a special example of an ACMC-$\infty$ coordinate system, where all multipole moments can be read off from the metric expansion.

Comparing with \eqref{eq: CH3: metric components to linear order g_00} and \eqref{eq: CH3: metric components to linear order g_0j} above, we have specified the constants $C_\l$ and $C_\l'$ such that the prefactors of the mass and current multipole moments are respectively $8/3\pi$ and $4/\pi$. This is the natural normalization such that $\g = M$ is the total mass in the spacetime and $\S_{12}, \S_{34}$ are the angular momenta (for a biaxially symmmetric spacetime) of the spacetime. This follows from the Newtonian expansion in arbitrary spacetime dimension $d$ as derived by Myers and Perry \cite{Myers:1986}; for example:
\begin{align*}
    g_{00}&\approx 1+\frac{16\pi}{(d-2)\Omega_{d-2}}\frac{M}{r^{d-3}}\,,
\end{align*}
which leads to the $8/(3\pi)$ prefactor in five dimensions that we used above. For convenience, we have also chosen the prefactors of the higher-order multipole moments to be the same as the leading order multipole.

The definition \eqref{eq: CH3: definition ACMC coordinates 5D} of ACMC-$N$ coordinates in five dimensions is our main new result, and represents an unambiguous definition of mass and current multipoles for five dimensional stationary, vacuum asymptotically flat solutions.

Note that it is natural to call $\g_{A_\l}$ the mass $\l$-pole so that the mass is the mass monopole (or $0$-pole), just as in four dimensions. Similarly, it is natural to call $\S_{bc A_{\l-1}}$ the current $\l$-pole. In particular, the current dipole is then $\S_{bc}$ --- i.e. the angular momentum tensor of the spacetime --- and is the leading order multipole appearing in the expansion of $g_{0j}$ (again, analogous to the four dimensional case). Note that in five dimensions, natural (geometric) units imply $[\g_{A_\l}]=L^{2+\l}=[\S_{bcA_{\l-1}}]$.

As mentioned above, a previous attempt had been made in \cite{Tanabe:2010} to define five-dimensional multipoles using the Geroch-Hansen formalism (albeit only for biaxially-symmetric spacetimes). They reported that octopoles and higher multipoles would be ambiguous and therefore ill-defined; however, in Appendix \ref{app:GH} we show this ambiguity is in fact not present. Even so, it is not immediately clear that the Geroch-Hansen formalism generalizes completely to five dimensions, as we also discuss in Appendix \ref{app:GH}.
In any case, the multipoles as defined here in the ACMC-$N$ formalism are entirely well-defined and unambiguous at all orders, which mean they are independent of the specific ACMC-$N$ coordinate system used. We show this explicitly in Section \ref{sec:proofinvariance}.

\subsection{Spacetimes with biaxial symmetry}\label{sec:biaxial}
In many applications, we are only interested in spacetimes that enjoy azimuthal symmetries. In four spatial dimensions, there are two independent azimuthal symmetries that can be preserved. Using Hopf coordinates \eqref{eq: CH3: definition Hopf coordinates}, whose definition we repeat here:
\begin{align}
    \label{eq:Hopfrepeated}
    &x_1=r\sin\theta\cos\phi_1\,,\,
    x_2=r\sin\theta\sin\phi_1\,,\,
    x_3=r\cos\theta\cos\phi_2\,,\,
    x_4=r\cos\theta\sin\phi_2\,,\\\nonumber
    &\theta\in [0,\tfrac{\pi}{2}]\,,\quad \phi_1,\phi_2\in[0,2\pi)\,.
\end{align}
the two azimuthal angles can be taken to be $\phi_1$ and $\phi_2$.

For a stationary spacetime that additionally enjoys the symmetries under rotations in $\phi_1$ and $\phi_2$, the general ACMC expansion simplifies considerably. For example, these symmetries immediately imply that we must have
\begin{equation}\label{eq:gAloddvanish}
    \g_{A_\l}=0\quad\text{if $\l$ is odd}\,,
\end{equation}
since terms $\g_{A_\l}N_{A_\l}$ for $\l$ odd can never be invariant under the azimuthal symmetries. In addition, from these symmetries it further follows that every tensor $\g_{A_\l}$ (for $\l$ even) must be completely determined by one single coefficient that we can call $M_{\l}$. In particular, we have:
\begin{equation}
    \label{eq: CH3: angular dependecy of mass multipole is legendre}
    \g_{A_\l}N_{A_\l}=(-1)^{\l/2}M_{\l}P_{\l/2}\left(\cos 2\theta\right)\quad \text{if $\l$ is even.}
\end{equation}
Similar arguments can be used to show that:
\begin{equation}
    \S_{abA_\l}=0 \quad \text{ if $\l$ is odd.}
\end{equation}
and that the tensors $\S_{abA_{\l-1}}=0$ (for $\l$ odd) are completely determined by a \emph{pair} of coefficients $S^{(1)}_{\l},S^{(2)}_{\l}$, which enter in the metric as:
\begin{equation*}
    \label{eq: CH3: -sin phi_2 g_03 + cos phi_2 g_04}
    -\sin\phi_2\varepsilon_{3bca_\l}\S_{bcA_{\l-1}}N_{A_\l}+\cos\phi_2\varepsilon_{4bca_\l}\S_{bcA_{\l-1}}N_{A_\l}=-2r\cos^2\theta (-1)^{\frac{\l-1}{2}}S^{(2)}_{\l}P_{\frac{\l-1}{2}}(\cos2\theta)\,.
\end{equation*}
So we see that, as in four-dimensions, the extra demand of axisymmetry/ies reduces the mass multipole tensor to a single number at each order. In four dimensions, the current multipole tensors also reduce to a single number at each order --- a single tower of multipole moments with the angular momentum as its lowest non-zero entry. In five dimensions, we have two numbers at each order --- two towers of multipole moments with the two independent angular momenta $S^{(1)}_1, S^{(2)}_1$ as their lowest non-trivial entries.

After demanding these two azimuthal symmetries and converting the ACMC expansion \eqref{eq: CH3: definition ACMC coordinates 5D} to Hopf coordinates using \eqref{eq:Hopfrepeated}, we get the following ACMC-$N$ expansion for biaxially symmetric stationary spacetimes:
\clearpage
\begin{mdframed}[frametitle={Biaxial symmetric ACMC-$N$ coordinates }]
\begin{align}
    \nonumber
    g_{00}&=-1+\frac{8 M}{3\pi r^2}+\sum_{\substack{\l=2\\ \text{$\l$ even}}}^{N}\frac{8}{3\pi\,r^{\l+2}}\Bigg((-1)^{\frac{\l}{2}}M_\l P_{\frac{\l}{2}}+\sum_{\substack{\l'=0\\ \text{$\l'$ even}}}^{\l-2}c_{\l\l'}^{(00)}P_{\frac{\l'}{2}}\Bigg)+(N\,\text{mod}\,2)\times \\\nonumber &\qquad\qquad\qquad\frac{8}{3\pi r^{N+3}}\Bigg((-1)^\frac{N+1}{2}M_{N+1}P_{\frac{N+1}{2}}+\sum_{\substack{\l'\neq N+1 \\ \text{$\l'$ even}}}c_{N+1\,\l'}^{(00)}P_{\frac{\l'}{2}}\Bigg)+\O\left(1/r^{N+4}\right)\,,\\[1em]\nonumber
    g_{0\phi_1}&=-2r\sin^2\theta\Bigg[\sum_{\substack{\l=1 \\ \text{$\l$ odd}}}^N\frac{4}{\pi r^{\l+2}}\Bigg((-1)^{\frac{\l-1}{2}}S^{(1)}_{\l}P_{\frac{\l-1}{2}}+\sum_{\substack{\l'=1 \\ \text{$\l'$ odd}}}^{\l-2}c_{\l\l'}^{(0\phi_1)}P_{\frac{\l'-1}{2}}\Bigg)+((N+1)\,\text{mod}\,2)\times\\ \nonumber
    &\qquad\qquad\qquad\frac{-4}{\pi r^{N+3}} \Bigg((-1)^\frac{N}{2}S^{(1)}_{N+1}P_{\frac{N}{2}}+\sum_{\substack{\l'\neq N+1 \\ \text{$\l'$ odd}}}c^{(0\phi_1)}_{N+1\,\l'}P_{\frac{\l'-1}{2}}\Bigg)+\O\left(1/r^{N+4}\right)\Bigg]\,,\\[1em]\nonumber
    g_{0\phi_2}&=-2r\cos^2\theta\Bigg[\sum_{\substack{\l=1 \\ \text{$\l$ odd}}}^N\frac{4}{\pi r^{\l+2}}\Bigg((-1)^{\frac{\l-1}{2}}S^{(2)}_{\l}P_{\frac{\l-1}{2}}+\sum_{\substack{\l'=1 \\ \text{$\l'$ odd}}}^{\l-2}c_{\l\l'}^{(0\phi_2)}P_{\frac{\l'-1}{2}}\Bigg)+((N+1)\,\text{mod}\,2)\times\\ \nonumber
    &\qquad\qquad\qquad\frac{-4}{\pi r^{N+3}} \Bigg((-1)^\frac{N}{2}S^{(1)}_{N+1}P_{\frac{N}{2}}+\sum_{\substack{\l'\neq N+1 \\ \text{$\l'$ odd}}}c^{(0\phi_2)}_{N+1\,\l'}P_{\frac{\l'-1}{2}}\Bigg)+\O\left(1/r^{N+4}\right)\Bigg]\,,\\\nonumber
    g_{rr}&=1+\sum_{\substack{\l=0 \\ \text{$\l$ even}}}^N\frac{1}{r^{\l+2}}\sum_{\substack{\l'=0 \\ \text{$\l'$ even}}}^{\l}c_{\l\l'}^{(rr)}P_{\frac{\l'}{2}}+\frac{N\,\text{mod}\,2}{r^{N+3}}\sum_{\text{$\l'$ even}}c_{N+1\,\l'}^{(rr)}P_{\frac{\l'}{2}}+\O\left(1/r^{N+4}\right)\,,\\ \nonumber
    g_{\theta\theta}&=r^2 \bigg[1+\sum_{\substack{\l=0 \\ \text{$\l$ even}}}^N\frac{1}{r^{\l+2}}\sum_{\substack{\l'=0 \\ \text{$\l'$ even}}}^{\l}c_{\l\l'}^{(\theta\theta)}P_{\frac{\l'}{2}}+\frac{N\,\text{mod}\,2}{r^{N+3}}\sum_{\text{$\l'$ even}}c_{N+1\,\l'}^{(\theta\theta)}P_{\frac{\l'}{2}}+\O\left(1/r^{N+4}\right)\bigg]\,,\\ \nonumber
    g_{\phi_1\phi_1}&=r^2\sin^2\theta\bigg[1+\sum_{\substack{\l=0 \\ \text{$\l$ even}}}^N\frac{1}{r^{\l+2}}\sum_{\substack{\l'=0 \\ \text{$\l'$ even}}}^{\l}c_{\l\l'}^{(\phi_1\phi_1)}P_{\frac{\l'}{2}}+\frac{N\,\text{mod}\,2}{r^{N+3}}\sum_{\text{$\l'$ even}}c_{N+1\,\l'}^{(\phi_1\phi_1)}P_{\frac{\l'}{2}}+\O\left(1/r^{N+4}\right)\bigg]\,,\\ \nonumber
    g_{\phi_2\phi_2}&=r^2\cos^2\theta\bigg[1+\sum_{\substack{\l=0 \\ \text{$\l$ even}}}^N\frac{1}{r^{\l+2}}\sum_{\substack{\l'=0 \\ \text{$\l'$ even}}}^{\l}c_{\l\l'}^{(\phi_2\phi_2)}P_{\frac{\l'}{2}}+\frac{N\,\text{mod}\,2}{r^{N+3}}\sum_{\text{$\l'$ even}}c_{N+1\,\l'}^{(\phi_2\phi_2)}P_{\frac{\l'}{2}}+\O\left(1/r^{N+4}\right)\bigg]\,,\\\nonumber
    g_{\phi_1\phi_2}&=r^2\cos^2\theta\sin^2\theta\bigg[\sum_{\substack{\l=0 \\ \text{$\l$ even}}}^N\frac{1}{r^{\l+2}}\sum_{\substack{\l'=0 \\ \text{$\l'$ even}}}^{\l}c_{\l\l'}^{(\phi_1\phi_2)}P_{\frac{\l'}{2}}+\frac{N\,\text{mod}\,2}{r^{N+3}}\sum_{\text{$\l'$ even}}c_{N+1\,\l'}^{(\phi_1\phi_2)}P_{\frac{\l'}{2}}+\O\left(1/r^{N+4}\right)\bigg]\,,\\\label{eq: CH3: ACMC conditions 5D axisymmetry}
    g_{r\theta}&=r\cos\theta\sin\theta\bigg[\sum_{\substack{\l=0 \\ \text{$\l$ even}}}^N\frac{1}{r^{\l+2}}\sum_{\substack{\l'=0 \\ \text{$\l'$ even}}}^{\l}c_{\l\l'}^{(r\theta)}P_{\frac{\l'}{2}}+\frac{N\,\text{mod}\,2}{r^{N+3}}\sum_{\text{$\l'$ 
    even}}c_{N+1\,\l'}^{(r\theta)}P_{\frac{\l'}{2}}+\O\left(1/r^{N+4}\right)\bigg]\,.
\end{align}
\end{mdframed}
All the $c^{(ij)}_{\l\l'}$'s are constants and in the Legendre polynomials we have suppressed the argument $\cos2\theta$. The factors $N\,\text{mod}\,2$ and $(N+1)\,\text{mod}\,2$ are present to emphasize that $M_{\l}$ and $S^{(i)}_{\l+1}$ are zero if $\l$ is odd, and to ensure that the order of the Legendre polynomials is a natural number. As in the general case, there may be multiple ACMC coordinate systems describing the same source. Depending on the coordinate system in which one works, the constants $c^{(ij)}_{\l\l'}$ may vary, but the multipole moments $M_\l$, $S^{(1)}_\l$ and $S^{(2)}_\l$ are the same in each coordinate system (see below in Section \ref{sec:proofinvariance}). Finally, note that the asymptotically Cartesian coordinate system \eqref{eq: CH3: ACMC conditions 5D axisymmetry} is automatically mass-centered in the case of biaxially symmetry, since $M_1=0$ from \eqref{eq:gAloddvanish}.

We have omitted the terms $g_{0r}$, $g_{0\theta}$, $g_{r\phi_1}$, $g_{r\phi_2}$, $g_{\theta\phi_1}$ and $g_{\theta\phi_2}$, since we are unaware of any example of a biaxially symmetric metric that contains them. However, to be complete, they are given by:
\begin{align*}
    g_{0r}&=\sum_{\substack{\l=1 \\ \text{$\l$ odd}}}^N\frac{1}{r^{\l+2}}\sum_{\substack{\l'=1 \\ \text{$\l'$ odd}}}^{\l}c_{\l\l'}^{(0 r)}P_{\frac{\l'-1}{2}}+\frac{(N+1)\,\text{mod}\,2}{r^{N+3}}\sum_{\text{$\l'$ 
    odd}}c_{N+1\,\l'}^{(0 r)}P_{\frac{\l'-1}{2}}+\O\left(1/r^{N+4}\right)\,,\\\nonumber
    g_{0\theta}&=r\bigg[\sum_{\substack{\l=1 \\ \text{$\l$ odd}}}^N\frac{1}{r^{\l+2}}\sum_{\substack{\l'=1 \\ \text{$\l'$ odd}}}^{\l}c_{\l\l'}^{(0 \theta)}P_{\frac{\l'-1}{2}}+\frac{(N+1)\,\text{mod}\,2}{r^{N+3}}\sum_{\text{$\l'$ 
    odd}}c_{N+1\,\l'}^{(0 \theta)}P_{\frac{\l'-1}{2}}+\O\left(1/r^{N+4}\right)\bigg]\,,\\\nonumber
    g_{r\phi_1}&=r\sin^2\theta\bigg[\sum_{\substack{\l=0 \\ \text{$\l$ even}}}^N\frac{1}{r^{\l+2}}\sum_{\substack{\l'=0 \\ \text{$\l'$ even}}}^{\l}c_{\l\l'}^{(r\phi_1)}P_{\frac{\l'}{2}}+\frac{N\,\text{mod}\,2}{r^{N+3}}\sum_{\text{$\l'$ 
    even}}c_{N+1\,\l'}^{(r\phi_1)}P_{\frac{\l'}{2}}+\O\left(1/r^{N+4}\right)\bigg]\,,\\\nonumber
    g_{r\phi_2}&=r\cos^2\theta\bigg[\sum_{\substack{\l=0 \\ \text{$\l$ even}}}^N\frac{1}{r^{\l+2}}\sum_{\substack{\l'=0 \\ \text{$\l'$ even}}}^{\l}c_{\l\l'}^{(r\phi_2)}P_{\frac{\l'}{2}}+\frac{N\,\text{mod}\,2}{r^{N+3}}\sum_{\text{$\l'$ 
    even}}c_{N+1\,\l'}^{(r\phi_2)}P_{\frac{\l'}{2}}+\O\left(1/r^{N+4}\right)\bigg]\,,\\\nonumber
    g_{\theta\phi_1}&=r^2\cos\theta\sin\theta\bigg[\sum_{\substack{\l=0 \\ \text{$\l$ even}}}^N\frac{1}{r^{\l+2}}\sum_{\substack{\l'=0 \\ \text{$\l'$ even}}}^{\l}c_{\l\l'}^{(\theta\phi_1)}P_{\frac{\l'}{2}}+\frac{N\,\text{mod}\,2}{r^{N+3}}\sum_{\text{$\l'$ 
    even}}c_{N+1\,\l'}^{(\theta\phi_1)}P_{\frac{\l'}{2}}+\O\left(1/r^{N+4}\right)\bigg]\,,\\\nonumber
    g_{\theta\phi_2}&=r^2\cos\theta\sin\theta\bigg[\sum_{\substack{\l=0 \\ \text{$\l$ even}}}^N\frac{1}{r^{\l+2}}\sum_{\substack{\l'=0 \\ \text{$\l'$ even}}}^{\l}c_{\l\l'}^{(\theta\phi_2)}P_{\frac{\l'}{2}}+\frac{N\,\text{mod}\,2}{r^{N+3}}\sum_{\text{$\l'$ 
    even}}c_{N+1\,\l'}^{(\theta\phi_2)}P_{\frac{\l'}{2}}+\O\left(1/r^{N+4}\right)\bigg]\,.
\end{align*}

\subsection{Invariance of multipole moments} \label{sec:proofinvariance}

We consider two ACMC-$N$ systems $\{x^{\alpha\prime}\}$ and $\{x^\alpha\}$ that satisfy \eqref{eq: CH3: definition ACMC coordinates 5D} which are related by the coordinate transformation
\begin{equation*}
    x^{\alpha\prime}=x^\alpha+f^\alpha(x^j)\,.
\end{equation*}
(Note that we do not assume biaxial symmetry as in the previous section.) Given that these two coordinate systems are both ACMC-$N$, we prove in this section that the multipole tensors $\g_{A_\l},\S_{bc A_{\l-1}}$for $\l\leq N$ appearing in the ACMC-$N$ expansion \eqref{eq: CH3: definition ACMC coordinates 5D} of both coordinate systems are the same, thus proving that mass and current multipole moments are indeed well-defined by the ACMC-$N$ condition \eqref{eq: CH3: definition ACMC coordinates 5D}.

First, note that the $f^\alpha$ cannot depend on $x^0$ since this would spoil stationarity. We can expand the transformation functions in powers of $1/r$:
\begin{align}
    \label{eq: f0 in powers of 1/r}
    f_0&=\sum_{n=-2}^\infty r^{-(n+2)}f_0^n(\theta,\phi_1,\phi_2)\\
    \label{eq: fj in powers of 1/r}
    f_j&=\sum_{n=-2}^\infty r^{-(n+2)}f_j^n(\theta,\phi_1,\phi_2)\,.
\end{align}
No terms with $n<-2$ can be allowed in this expansion, since then one or both coordinate systems would not be asymptotically flat. Define now
\begin{equation}
    \label{eq: CH3: definition h_alphabeta in proof of invariance}
    h_{\alpha\beta}(x^{j\prime})=g_{\alpha'\beta'}(x^{j\prime})-\eta_{\alpha\beta}
\end{equation}
and expand $h_{\alpha\beta}(x^j)$ in powers of $1/r$:
\begin{equation}
    \label{eq: halphabeta in powers of 1/r}
    h_{\alpha\beta}(x^j)=\sum_{n=0}^\infty r^{-(n+2)}h_{\alpha\beta}^n(\theta,\phi_1,\phi_2)\,.
\end{equation}
Since $\{x^{\alpha\prime}\}$ is ACMC-$N$, we can compare with \eqref{eq: CH3: definition ACMC coordinates 5D} to see that we have no terms with $n<0$ and:
\begin{itemize}
    \item $h_{00}^0$ and $h^0_{jk}$ contain only monopoles\,, and $h_{0j}^0=0$\,,
    \item $h_{0j}^1$ and $h^1_{jk}$ contain only monopoles and dipoles\,, $h^1_{00}$ contains only monopoles\,,
    \item $h_{\alpha\beta}^n$ contains only poles of order $\l\leq n$ for $2\leq n \leq N$\,.
\end{itemize}
The metric tensors in the two coordinate systems are related by
\begin{equation*}
    g_{\mu\nu}(x)=\frac{\partial x^{\alpha'}}{\partial x^\mu}\frac{\partial x^{\beta'}}{\partial x^\nu}g_{\alpha'\beta'}\left(x'(x)\right)\,.
\end{equation*}
Taylor expanding $g_{\alpha'\beta'}\left(x'(x)\right)$ via \eqref{eq: CH3: definition h_alphabeta in proof of invariance} and writing this expression explicitly, we obtain
\begin{align}
    \label{eq: metric of first system as function of a and b}
    g_{\mu\nu}(x)=\eta_{\mu\nu}+h_{\mu\nu}(x)+a_{\mu\nu}(x)+b_{\mu\nu}(x)\,,
\end{align}
where
\begin{align}
    \label{eq: amunu definition}
    a_{\mu\nu}(x)&=(\delta_\mu^\alpha+f^{\alpha}_{\ \,,\,\mu})(\delta_\nu^\beta+f^{\beta}_{\ \,,\,\nu})\sum_{n=0}^\infty \left((n!)^{-1}f_{j_1}\dots f_{j_n}h_{\alpha\beta}(x)_{\,,\,j_1\dots j_n}\right)-h_{\mu\nu}(x)+f^{\alpha}_{\ \,,\,\mu}f_{\alpha\,,\,\nu}\,,\\
    \label{eq: bmunu definition}
    b_{\mu\nu}(x)&=f_{\mu\,,\,\nu}+f_{\nu\,,\,\mu}\,.
\end{align}
Expanding the above three expressions in powers of $1/r$ gives
\begin{align}
    \label{eq: gmunu in powers of 1/r}
    g_{\mu\nu}&=\eta_{\mu\nu}+\sum_{n=0}^\infty r^{-(n+2)}g^n_{\mu\nu}(\theta,\phi_1,\phi_2)\,,\\
    \label{eq: amunu in powers of 1/r}
    a_{\mu\nu}&=\sum_{n=-1}^\infty r^{-(n+2)}a^n_{\mu\nu}(\theta,\phi_1,\phi_2)\,,\\
    \label{eq: bmunu in powers of 1/r}
    b_{\mu\nu}&=\sum_{n=-1}^\infty r^{-(n+2)} b^n_{\mu\nu}(\theta,\phi_1,\phi_2)\,.
\end{align}
Then, demanding that the $f^\alpha$ are coordinate transformations between two ACMC-$N$ systems is equivalent to demanding that
\begin{itemize}
    \item $a^{-1}_{\mu\nu}+b^{-1}_{\mu\nu}=0$\,,
    \item $a_{00}^0+b_{00}^0$ and $a^0_{jk}+b^0_{jk}$ contain only monopoles\,, $a_{0j}^0+b_{0j}^0=0$\,,
    \item $a_{0j}^1+b_{0j}^1$ and $a^1_{jk}+b^1_{jk}$ contain only monopoles and dipoles\,, $a^1_{00}+b^1_{00}$ contains only monopoles\,,
    \item $a_{\alpha\beta}^n+b^n_{\alpha\beta}$ contains only poles of order $\l\leq n$ for $2\leq n \leq N$\,.
\end{itemize}

We can now start showing that the multipole moments $\g_{A_\l}$ and $\S_{bcA_{\l-1}}$ in both ACMC-$N$ systems are equal for $\l\leq N$. Let us start with computing
\begin{align}
    \nonumber
    f_{\mu\,,\,j}&=\sum_{n=-2}^\infty \partial_j\left(r^{-(n+2)}f_{\mu}^n(\theta,\phi_1,\phi_2)\right)\\\nonumber
    &=\sum_{n=-2}^\infty\left(-(n+2)r^{-(n+3)}n_j f_\mu^n(\theta,\phi_1,\phi_2)+r^{-(n+2)}f_\mu^n(\theta,\phi_1,\phi_2)_{\,,\,j}\right)\\
    \label{eq: fmu,nu in powers of 1/r}
    &=\sum_{n=-2}^\infty\left(r^{-(n+3)}\left(-(n+2)n_j f_{\mu}^n(\theta,\phi_1,\phi_2)+rf_{\mu}^n(\theta,\phi_1,\phi_2)_{\,,\,j}\right)\right)\,.
\end{align}
Since the derivative on the spherical coordinates introduces a factor of $1/r$, the term in the inner brackets really is independent of $r$. Since $h_{\alpha\beta}(x)=\O(r^{-2})$ and $f_{\mu\,,\,j}=\O(r^{-1})$, the lowest order term of $a_{\mu\nu}$ is that of $f_{\ \,,\,i}^\alpha f_{\alpha\,,\,j}$, which is $\O(r^{-2})$. This implies
\begin{equation}
\label{eq: a^-1=0}
a^{-1}_{\mu\nu}=0\,,
\end{equation}
and therefore also
\begin{equation}
    \label{eq: b^-1=0}
    b^{-1}_{\mu\nu}=0\,.
\end{equation}
The lowest order term of $a_{\mu\nu}$ is then $r^{-2} a^0_{\mu\nu}$, where
\begin{equation}
    \label{eq: amunu^0= f^-2 f^-2}
    a^{0}_{\mu\nu}=(r f_{\alpha\,,\,\mu}^{-2})(r f_{\alpha\,,\,\nu}^{-2})\,.
\end{equation}
Applying this on $(\mu,\nu)=(0,0)$, we have, due to stationarity, that $a^0_{00}=0$. For the same reasons, we also have $b_{00}^0=0$. So $a^0_{00}+b^0_{00}=0$ and from \eqref{eq: metric of first system as function of a and b} it then follows that both systems have the same mass monopole moment. 

We move on to the current dipole moment. Consider
\begin{equation}
    \label{eq: a_0j}
    a_{0j}(x)=\sum_{n=1}^\infty (n!)^{-1}f_{j_1}\dots f_{j_n} h_{0j\,,\,j_1\dots j_n}+f^\alpha_{\ \, ,j}\sum_{n=0}^\infty (n!)^{-1}f_{j_1}\dots f_{j_n}h_{0\alpha\,,\,j_1\dots j_n}\,.
\end{equation}
To find $a^1_{0j}$, we look at $r^{-3}$ terms. The $r^{-3}$ term of the first term in \eqref{eq: a_0j} is given by $f_i^{-2}\partial_i(r^{-2}h_{0j}^0)$ which is zero since $h_{0j}^0=0$. For the second term in \eqref{eq: a_0j}, first consider $a^0_{0j}$. By invoking \eqref{eq: amunu^0= f^-2 f^-2} on $(\mu,\nu)=(0,j)$, we see that 
\begin{equation}
    \label{eq: a^0_0j=0}
    a^0_{0j}=0\,.
\end{equation}
Using \eqref{eq: b^-1=0} and \eqref{eq: fmu,nu in powers of 1/r}, we obtain
\begin{equation}
    \label{eq: f^-2_0j=0}
    0=r^{-1}b^{-1}_{0j}=f^{-2}_{0\,,\,j}\,.
\end{equation}
Now return to the $r^{-3}$ term of the second term of \eqref{eq: a_0j}. Since $h_{0\alpha}=\O(r^{-2})$ and $f_{\alpha,j}=\O(r^{-1})\,$, the coefficient of this term is given by
\begin{equation*}
rf^{-2}_{0\,,\,j}h_{00}^0+rf^{-2}_{i\,,\,j}h_{0i}^0=rf^{-2}_{0\,,\,j}h_{00}^0=0\,,
\end{equation*}
due to \eqref{eq: f^-2_0j=0}. So altogether, we have
\begin{equation}
    \label{eq: a^1_0j=0}
    a^1_{0j}=0\,.
\end{equation}
To find $b^1_{0j}$, we must look at the $r^{-3}$ term of \eqref{eq: fmu,nu in powers of 1/r} with $(\mu,\nu)=(0,j)$. We find
\begin{equation*}
a^1_{0j}+b^1_{0j}=b^1_{0j}=-2n_jf_0^0+rf_{0\,,\,j}^0=r^3\partial_j\left(r^{-2}f_0^0\right)
\end{equation*}
Such a term is not of the form $\varepsilon_{jabc}\T_{ab}n_c$ for some tensor $\T_{A_2}$ hence it is a $1$-pole of the form $\B n_j$ or $\C_{ja}n_a $. Therefore, both systems have the same current dipole moment $\S_{ab}$.

Now we turn to the general case of $a^n_{\mu\nu}$ and $b^n_{\mu\nu}$ with $2\leq n \leq N$. First notice that, by virtue of \eqref{eq: b^-1=0}, we have
\begin{equation*}
    f_{0\,,\,j}^{-2}=0\,,\quad\text{and}\quad f_{(i\,,\,j)}^{-2}=0\,.
\end{equation*}
This shows that $f^{-2}_{\alpha}$ are pure monopole terms.
For $n\geq -1$, we prove by induction on $n\leq N+1$ that $f^{n-1}_\alpha$ contains only poles of order $\l\leq n$. We start with the base case.

Since $f^{-2}_j$ is monopole and due to \eqref{eq: amunu^0= f^-2 f^-2}, we see that
\begin{equation*}
    a_{jk}^0=(r f^{-2}_{\alpha\,,\,j})(rf_{\alpha\,,\,k}^{-2})=0\,.
\end{equation*}
Since $b^0_{jk}$ is given by
\begin{equation*}
    b^0_{jk}=r^2 \partial_{(k}\left(r^{-1}f^{-1}_{j)}\right)\,,
\end{equation*}
the condition \{$a_{jk}^0+b^0_{jk}$ contains only monopoles\} together with the fact that spatial derivatives do not change multipole order then imply that $f^{-1}_j$ contains only monopoles. Furthermore, $a_{0j}^0=0$ and 
\begin{equation*}
    b^0_{0j}=r^2\partial_j\left(r^{-1}f_{0}^{-1}\right)\,,
\end{equation*}
hence the condition \{$a^0_{0j}+b^0_{0j}=0$\} implies that $f_0^{-1}$ contains only monopoles. Thus, the base case is proven. 

Per our induction hypothesis, suppose that $f^{n'-1}_\alpha$ contains only poles of order $\l\leq n'$ for all $0<n'< n\leq N+1$. We prove that $f^{n-1}_\alpha$ contains only poles of order $\l\leq n$. Upon using \eqref{eq: amunu definition}, \eqref{eq: amunu in powers of 1/r}, \eqref{eq: f0 in powers of 1/r}, \eqref{eq: fj in powers of 1/r} and \eqref{eq: fmu,nu in powers of 1/r}, and the fact that spatial derivatives do not change multipole order, we can find the possible type of terms that $a^n_{\mu\nu}$ contains. The term $f^\alpha_{\ \,,\,\mu}f_{\alpha\,,\,\nu}$ in $a_{\mu\nu}$ gives the following terms in $a^n_{\mu\nu}$:
\begin{equation}
    \label{eq: first type of term in amunu^n}
    f_{\mu_1}^{p_1}f_{\mu_2}^{p_2}\quad\text{or}\quad f_{\mu_1}^{p_1}f_{\mu_2\,,\,j}^{p_2}\quad\text{or}\quad f_{\mu_1\,,\,j_1}^{p_1}f_{\mu_2\,,\,j_2}^{p_2}\quad\text{ with }p_1+p_2+4=n\,,
\end{equation}
where the counting is coming from $(p_1+3)+(p_2+3)=(n+2)$. The other term in $a_{\mu\nu}$ gives
\begin{equation}
\label{eq: second type of term in amunu^n}
f_{\mu_1\,,\,j_1}^{p_1}f_{\mu_2\,,\,j_2}^{p_2}f_{\mu_3}^{p_3}\dots f_{\mu_k}^{p_k} h_{\alpha\beta}^{q}\quad\text{ with }p_1+p_2+p_3+\dots+p_k+3k+q=n\,,
\end{equation}
with the counting coming from
$(p_1+3)+(p_2+3)+(p_3+2)+\dots+(p_k+2)+(q+2+k-2)=(n+2)$. Note that we must have $k\geq 1$. It is also possible that the first two factors of \eqref{eq: second type of term in amunu^n} are $f_{\mu_1\,,\, j_1}^{p_1}f_{\mu_2}^{p_2}$ or $f_{\mu_1}^{p_1}f_{\mu_2}^{p_2}$. Although obtained slightly differently, the result of the counting is still $p_1+p_2+p_3+\dots+p_k+3k+q=n$.\\  If $\l_1$ and $\l_2$ are the multipole order of respectively the first and second factor in \eqref{eq: first type of term in amunu^n}, then the multipole order of the product is $\l\leq \l_1+\l_2$. In principle if $p_i=-2$ then $\l_i\leq p_i+2$ but since $f^{-2}_{\alpha}$ is monopole, i.e. $\l=0$, they do not increase the multipole order, so henceforth we assume $p_i\geq -1$. Then due to \eqref{eq: first type of term in amunu^n} we have $p_1,p_2\leq n-3$ so we may use the induction hypothesis to obtain
\begin{equation*}
    \l\leq \l_1+\l_2\leq p_1+1+p_2+1= n-2
\end{equation*}
For the second type of term, let $L$ be the multipole order of $h^q_{\alpha\beta}$. The ACMC-$N$ conditions restrict $L\leq q$. Since $k\geq 1$ and $p_i \geq -1$, we have $p_i \leq n-3$. So by induction, the multipole order of \eqref{eq: second type of term in amunu^n} is given by
\begin{equation*}
    \l\leq \l_1+\dots +\l_k+L\leq (p_1+1)+\dots+(p_k+1)+q=p_1+\dots+p_k+k+q=n-2k\leq n-2\,,
\end{equation*}
since $k\geq 1$. This implies that $a^n_{\mu\nu}$ contains only poles of order $\l\leq n-2$.

Now we turn to $b^n_{\mu\nu}$, which is given by
\begin{align}
    \label{eq: b_00^n=0}
    b_{00}^n&=0\,,\\\nonumber
    b_{0j}^n&=-(n+1) n_j f^{n-1}_0+ rf_{0\,,\,j}^{n-1}=r^{n+2}\partial_j\left(r^{-(n+1)}f_0^{n-1}\right)\,,\\\nonumber
    b_{jk}^n&=2r^{n+2}\partial_{(j}\left(r^{-(n+1)}f_{k)}^{n-1}\right)\,.
\end{align}
The ACMC-$N$ conditions together with the fact that $a^n_{\mu\nu}$ contains only poles of order $\l\leq n-2$ imply that $f^{n-1}_\alpha$ contains only poles of order $\l\leq n$, which proves by induction our assertion. Furthermore, because of \eqref{eq: b_00^n=0} and since the vector $b_{0j}^n$ is not of the form $\varepsilon_{jbca_n}\T_{bcA_{n-1}}N_{A_n}$, it follows from \eqref{eq: metric of first system as function of a and b} that the two coordinate systems have the same mass $n$-pole and the same current $n$-pole. Moreover, for $n=N+1$ we still have that $f^{N+1}_\alpha$ contains only poles of order $\l\leq N+1$ and that $a^{N+1}_{\alpha\beta}$ contains only poles of order $\l\leq N-1$. Thus, although the ACMC conditions no longer hold for $n=N+1$, the mass and current $(N+1)$-poles of the two coordinate systems are the same.

\medskip

Thorne provided a similar proof of the well-definedness of the ACMC-$N$ multipole moments in four dimensions in his original paper \cite{Thorne:1980}. Our proof in five dimensions follows the essence of this proof --- expanding in powers of $1/r$ and proving by induction that there is an upper bound for the order of angular dependency of the transformation functions $f_\alpha$. The main difference of our five-dimensional and Thorne's four-dimensional proof is that we mainly use the STF-decomposition of angular dependencies instead of decomposing in (scalar and vector) spherical harmonics. Similarly, we do not use parity to prove the invariance of current multipole moments, choosing again instead to use properties of the STF (and ASTF) decomposition.

\section{Examples}\label{sec:examples}
In this Section, we use our formalism to calculate the mass and current multipole moments for the Myers-Perry black hole, black rings with one or two angular momenta, and smooth supersymmetric multicentered geometries.

\subsection{Myers-Perry black hole}
The metric for the 5-dimensional MP black hole with mass parameter $m$ and rotation parameters $a,b$ is given by \cite{Myers:1986,myers:2011}
\begin{align}
    \nonumber
    ds^2&=-dt^2+\frac{\rho^2}{\Delta}dr^2+(a^2-b^2)\sin^4\theta\,d\phi_1^2+(b^2-a^2)\cos^4\theta \,d\phi_2^2\\\label{eq: metric Myers-Perry}
    &\quad +\frac{2m}{\rho^2}(dt-a\sin^2\theta \,d\phi_1-b\cos^2\theta \,d\phi_2)^2+\rho^2(d\theta^2+\sin^2\theta\, d\phi_1^2+\cos^2\theta\, d\phi_2^2)\,,
\end{align}
where
\begin{align*}
    \rho^2=r^2+a^2\cos^2\theta+b^2\sin^2\theta\,,\quad \Delta=\frac{(r^2+a^2)(r^2+b^2)}{r^2}-2m\,.
\end{align*}
If we denote the mass of the black hole by $M$ and the angular momenta in the $x_1x_2$-plane and the $x_3x_4$-plane by respectively $J^{(1)}$ and $J^{(2)}$, then the physical quantities $M$, $\Jone$ and $\Jtwo$ are related to the MP parameters via
\begin{equation}\label{eq:MJforMP}
    M=\frac{6\pi}{8}m\,,\quad \Jone=\frac{\pi}{2}ma=\frac{2}{3}Ma\,,\quad \Jtwo=\frac{\pi}{2}mb=\frac{2}{3}Mb\,.
\end{equation}

Although it is perhaps not immediately obvious from how it is written, the metric (\ref{eq: metric Myers-Perry}) is ACMC-0, meaning the leading order asymptotic expansion is indeed the flat space expression \eqref{eq:flatspace}. However, the metric is not ACMC to a higher order, as can be seen from e.g. the expansion of $g_{rr}$:
\be g_{rr}=1+\frac{2m-a^2-b^2+a^2\cos^2\theta+b^2\sin^2\theta}{r^2} + \mathcal{O}(r^{-4}),\ee
which contains a term $P_1(\cos2\theta)$ at order $r^{-2}$, which implies $N<1$ in the ACMC-$N$ expansion \eqref{eq: CH3: ACMC conditions 5D axisymmetry}.

A suitable coordinate transformation is needed to bring the metric to ACMC-$N$ form with $N>0$. The general transformation needed is of the form:
\begin{equation}\label{eq:coordtransfrrp}
    r=r'+\sum_{\substack{n=1\\n\text{ odd}}}^m\frac{F_n(\theta')}{(r')^n}\,,\quad \theta=\theta'+\sum_{\substack{n=2\\n\text{ even}}}^m\frac{G_n(\theta')}{(r')^n}\,,
\end{equation}
with the (isometry) coordinates $t,\phi_1,\phi_2$ unaltered, and with:
\begin{equation}
    \label{eq: CH4: F_n(theta')}
    F_n(\theta')=\sum_{\substack{k=3\\ k\text{ odd}}}^{n+2} u_k^{(n)} P_{\frac{k-1}{2}}(\cos 2\theta')\,, \qquad G_n(\theta')=\cos\theta'\sin\theta'\sum_{\substack{k=0\\ k\text{ even}}}^{n-2}v_k^{(n)}P_{\frac{k}{2}}(\cos2\theta')\,,
\end{equation}
The upper limit $m$ in \eqref{eq:coordtransfrrp} and the coefficients $u_k^{(n)},v_k^{(n)}$ in \eqref{eq: CH4: F_n(theta')} should then be chosen such that the resulting metric is ACMC-$N$ to the specified order $N$.

For example, in the case at hand, the transformation:
\begin{align*}    r&=r'+\frac{(a^2-b^2)\cos2\theta'}{4r'}\,,\qquad \theta=\theta'+\frac{(a^2+b^2)\cos\theta'\sin\theta'}{2(r')^{ 2}}\,
\end{align*}
is sufficient to render the metric ACMC-$1$ in the coordinates $(t,r',\theta',\phi_1,\phi_2)$. Repeating this procedure to find an ACMC-$N$ metric expansion for arbitrarily high $N$, we find that the multipoles of the Myers-Perry black hole are given by:
\begin{equation*}\label{eq:MPBHmultipoles}
    M_{2n}=(a^2-b^2)^{n}M\,,\quad S_{2n+1}^{(1)}=\frac{2n+1}{n+1}(a^2-b^2)^nJ^{(1)}\,,\quad S_{2n+1}^{(2)}=\frac{2n+1}{n+1}(a^2-b^2)^nJ^{(2)}\,,
\end{equation*}
for all $n\geq 0$, and where we used \eqref{eq:MJforMP}.
These coefficients can be compared to the multipole moments of four-dimensional Kerr: $M_{2n} = M(-a^2)^n$ and $S_{2n+1} = Ma(-a^2)^n$. Perhaps the most striking difference between the Myers-Perry and Kerr multipole towers is that the Myers-Perry the mass and current multipole moments can all vanish (except the mass and angular momenta themselves) in the case of equal angular momentum sizes, $|a|=|b|$. Interestingly, in four-dimensions, a similar phenomenon is also possible (although clearly not for Kerr): a particular four-dimensional almost-BPS (extremal) black hole was found in \cite{Bah:2021jno} for which $M,J$ were non-zero but all other higher-order multipoles vanished identically. In \cite{Bah:2021jno} such black holes were called \emph{purest-spinning black holes} and appear to be a special case in four dimensions; by contrast, for five dimensional black holes,  vanishing higher mass and current multipole moments simply requires two equal angular momenta.

\subsection{Black rings}\label{sec:BRmultipoles}

We will discuss the mass and current multipoles of black rings in this Section, first focusing on the simple case of black rings with one angular momentum before moving on to the more general two-angular momenta case.

\subsubsection{One angular momentum}
The metric of a black ring rotating in the $\phi_1$-direction is determined by (dimensionless) parameters $0<\nu\leq\lambda<1$ and a dimensionful radius $R$, and is given in so-called ring coordinates as  \cite{Emparan:2006}:
\begin{align}
    \nonumber
    ds^2&=-\frac{A(y)}{A(x)}\left(dt-CR\frac{1+y}{A(y)}d\tilde{\phi}_1\right)^2\\\label{eq: CH4: metric black ring}
    &\quad+\frac{R^2}{(x-y)^2}A(x)\left(-\frac{B(y)}{A(y)}d\tilde\phi_1^2-\frac{dy^2}{B(y)}+\frac{dx^2}{B(x)}+\frac{B(x)}{A(x)}d\tilde\phi_2^2\right)\,,
\end{align}
where $A$ and $B$ are functions and $C$ a constant defined by
\begin{equation*}
    A(\xi)=1+\lambda\xi\,,\quad B(\xi)=(1-\xi^2)(1+\nu\xi)\,,\quad C=\sqrt{\lambda(\lambda-\nu)\frac{1+\lambda}{1-\lambda}}\,.
\end{equation*}
The coordinates are related to asymptotically flat Hopf coordinates as:
\begin{equation}
    \label{eq: relation between ring coordinates and asymptotically flat hopf}
    x=-\frac{r^2-2\Tilde{R}^2\cos^2\theta}{r^2}\,,\quad y=-\frac{r^2+2\Tilde{R}^2\sin^2\theta}{r^2}\,,\quad \Tilde{R}=R\frac{1-\lambda}{1-\nu}\,,\quad ({\phi}_1,{\phi}_2)=\frac{1-\nu}{\sqrt{1-\lambda}}(\tilde\phi_1,\tilde\phi_2)\,.
\end{equation}

The two parameters $\nu$ and $\lambda$ are related to the shape and the rotational velocity of the ring. In particular, $\nu$ is proportional to the thickness of the ring while $\lambda/\nu$ determines the rotational velocity of the ring. To avoid singularities at $x=-1$, $y=-1$ and $x=1$ we must set
\begin{equation}
    \label{eq: CH4: black ring balance condition}
    \lambda=\frac{2\nu}{1+\nu^2}\,.
\end{equation}
Physically, this condition can be interpreted as tuning the centrifugal force to precisely counteract the self-attraction of the ring, for a given fixed ring mass and radius.
 Note that there exists a generalization \cite{pomeransky:2006} of \eqref{eq: CH4: metric black ring} that includes rotations in the $\phi_2$-direction, but for simplicity we do not discuss this here.

The resulting metric expressed in coordinates $(t,r,\theta,\phi_1,\phi_2)$ is ACMC-0, and we can use the algorithm detailed above to find further coordinate transformations to ACMC-$N$ systems for arbitrary $N>0$. Using \eqref{eq: CH4: black ring balance condition}, the mass and angular momentum are given by:
\begin{equation}
    \label{eq: CH4: mass and angular momentum black ring}
    M_0=\frac{3\pi R^2}{2}\frac{\nu}{(1-\nu)(\nu^2+1)}=M\,,\quad S_1^{(1)}=\frac{\pi R^3}{2}\nu\sqrt{\frac{2(\nu+1)^3}{(\nu^2+1)^3(1-\nu)^3}}=J^{(1)}\,,
\end{equation}
the first higher-order mass multipoles are given by:
\begin{align*}
    M_2&=M\left(\frac{M}{3\pi}\right)\frac{1+3\nu^2}{\nu}\,,\\
    M_4&=M\left(\frac{M}{3\pi}\right)^2\frac{3-\nu+21\nu^3+28\nu^4}{3\nu^2}\,,\\
    M_6&=M\left(\frac{M}{3\pi}\right)^3\frac{25-20\nu+293\nu^2-144\nu^3+899\nu^4-188\nu^5+735\nu^6}{25\nu^3}\,,\\
    M_8&=M\left(\frac{M}{3\pi}\right)^4\frac{1}{1225\nu^4}\Big(1225-1645\nu+20957\nu^2-19561\nu^3+107707\nu^4\\
    &\qquad\qquad\qquad\qquad\qquad-60135\nu^5+199663\nu^6-48771\nu^7+114160\nu^8\Big)\,,
\end{align*}
and the first few current multipole moments are given by
\begin{align*}
    S^{(1)}_3&=J^{(1)}R^2\left(\frac{2}{\pi}\right)\frac{3(1+3\nu^2)}{4(1-\nu)(1+\nu^2)}\,\\
    S_5^{(1)}&=J^{(1)}R^4\left(\frac{2}{\pi}\right)\frac{5-\nu+33\nu^2-3\nu^3+46\nu^4}{3(1-\nu)^{2}(1+\nu^2)^2}\,\\
    S^{(1)}_7&=J^{(1)}R^6\left(\frac{2}{\pi}\right)\frac{175-90\nu+1869\nu^2-612\nu^3+5697\nu^4-834\nu^5+6995\nu^6}{100(1-\nu)^3(1+\nu^2)^3}\,\\
    S^{(1)}_9&=J^{(1)}R^8\left(\frac{2}{\pi}\right)\frac{3}{1225(1-\nu)^4(1+\nu^2)^4}\Big(735-665\nu+11183\nu^2-7289\nu^3+55133\nu^4\\
    &\quad\qquad\qquad\qquad\qquad\qquad\qquad-22275\nu^5+104957\nu^6-19179\nu^7+65560\nu^8\Big)\,.
\end{align*}
The multipole moments do not appear to be able to be written in closed form as the Myers-Perry multipole moments could. We do observe the following relations for the (dimensionless) multipole moment expressions:
\begin{align*}
    \frac{M_{2n}}{M^{n+1}}&=\nu^{-n}\times\text{ ($2n$-th degree polynomial in $\nu$),}\\
    \frac{S^{(1)}_{2n+1}}{J^{(1)}R^{2n}}&=(1-\nu)^{-n}(1+\nu^2)^{-n}\times\text{ ($2n$th degree polynomial in $\nu$),}
\end{align*}
for all $n\geq 0$.  The current multipoles can also be written as
\begin{equation*}
    \frac{S^{(1)}_{2n+1}}{J^{(1)}M^n}=\nu^{-n}\times\text{ ($2n$th degree polynomial in $\nu$).}
\end{equation*}
Note also that the multipole moments reduce correctly to the Myers-Perry black hole multipole moments \eqref{eq:MPBHmultipoles} in the Myers-Perry limit $\nu\to 1$ and $R\to 0$.

The multipole moments can be used to easily explicitly demonstrate non-uniqueness of stationary vacuum solutions in five dimensions. In the window:
\be \frac{27}{32} < \frac{27\pi}{32}\frac{(J^{(1)})^2}{M^3} = \frac{(1+\nu)^3}{8\nu} < 1,\ee
there exist three different objects with the same $M$ and $J^{(1)}$ \cite{Emparan:2006}: the Myers-Perry black hole, the \emph{fat} black ring with $0<\nu<1/2$, and the \emph{thin} black ring with $1/2<\nu<1$. All three of these objects have, for example, different mass quadrupoles. As an explicit example, consider $J=0.595M^{3/2}$. The fat, resp. thin black rings have $\nu=0.279$, resp. $\nu=0.865$. The Myers-Perry black hole, thin black ring, and fat black ring with the same mass and angular momentum $J=0.595M^{3/2}$ then have mass quadrupoles given by: $M_2^\text{MP}=0.796M^2, M_2^\text{thin}= 0.470M^2, M_2^\text{fat}=0.398M^2$.

\subsubsection{Two angular momenta}
The metric of a black ring rotating in both the $\phi_1$- and $\phi_2$-direction is determined by the dimensionless parameters $0\leq \alpha < 1$ and $2\sqrt{\alpha}\leq \nu <1+\alpha$ and the dimensionful parameter $k$, which sets the scale of the black ring; we will use $R=k\sqrt{2(1+\nu^2)}$ instead as $R$ then  agrees with the one-angular momentum black-ring radius $R$ (i.e. when $\alpha=0$). In ring coordinates it is given by \cite{pomeransky:2006}:
\begin{align*}
    ds^2&=\frac{H(y,x)}{H(x,y)}(dt+\Omega)^2+\frac{F(x,y)}{H(y,x)}d\phi_1^2+2\frac{J(x,y)}{H(y,x)}d\phi_1d\phi_2-\frac{F(y,x)}{H(y,x)}d\phi_2^2\\
    &\quad-\frac{2k^2}{(x-y)^2(1-\alpha)^2}\left(\frac{dx^2}{G(x)}-\frac{dy^2}{G(y)}\right)\,,
\end{align*}
where $\Omega$ is given by 
\begin{align*}
    \Omega&=-\frac{2k\nu\sqrt{(1+\alpha)^2-\nu^2}}{H(y,x)}\bigg((1-x^2)y\sqrt{\alpha}d\phi_2+\frac{(1+y)}{(1-\nu+\alpha)}\\
    &\quad\big(1+\nu-\alpha+x^2y\alpha(1-\nu-\alpha)+2\alpha x(1-y)\big)d\phi_1\bigg)\,,
\end{align*}
and the functions $G$, $H$, $J$, and $F$ are defined as
\begin{align*}
    G(x)&=(1-x^2)(1+\nu x+\alpha x^2)\,,\\
    H(x,y)&=1+\nu^2-\alpha^2+2\alpha\nu(1-x^2)y+2x\nu(1-y^2\alpha^2)+x^2y^2\alpha(1-\nu^2-\alpha^2)\,,\\
    J(x,y)&=\frac{2k^2(1-x^2)(1-y^2)\nu\sqrt{\alpha}}{(x-y)(1-\alpha)^2}\big(1+\nu^2-\alpha^2+2(x+y)\nu\alpha-xy\alpha(1-\nu^2-\alpha^2)\big)\,,\\
    F(x,y)&=\frac{2k^2}{(x-y)^2(1-\alpha)^2}\Big(G(x)(1-y^2)\left(\left(\left(1-\alpha\right)^2-\nu^2\right)(1+\alpha)+y\nu(1-\nu^2+2\alpha-3\alpha^2)\right)\\
    &\quad+G(y)\big(2\nu^2+x\nu\left((1-\alpha)^2+\nu^2\right)+x^2\left((1-\alpha)^2-\nu^2\right)(1+\alpha) \\
    &\quad+x^3\nu(1-\nu^2-3\alpha^2+2\alpha^3)+x^4(1-\alpha)\alpha(-1+\nu^2+\alpha^2)\big)\Big)\,.
\end{align*}
Asymptotically flat Hopf coordinates are obtained by the transformation \eqref{eq: relation between ring coordinates and asymptotically flat hopf} plus an additional rescaling of the radial coordinate:
\begin{equation*}
    r\to \left(-\frac{\left(\nu ^2+1\right)^2 (\alpha -\nu +1)}{(\alpha -1) (\alpha +1)^2}\right)^{-\frac{1}{2}} r\,.
\end{equation*}
The first few mass multipole moments are then found to be
\begin{align*}
    M&=\frac{3\pi R^2}{2}\frac{ \nu  }{(1+\nu^2) (\alpha -\nu +1)}\,,\\
    M_2&=\frac{2M}{3\pi}\frac{R^2 \left(\alpha ^3-5 \alpha ^2+\alpha  \left(3 \nu ^2-8 \nu -5\right)+3 \nu ^2+1\right)}{(\alpha -1)^2 (1+\nu^2) (\alpha -\nu +1)}\,,\\
    M_4&=\frac{2M}{3\pi}\frac{R^4}{3 (\alpha -1)^4 (1+\nu^2)^2 (\alpha -\nu +1)^2} \Big(3 \alpha ^4 \left(7 \nu ^2-11 \nu -1\right)+\alpha ^3 \left(-3 \nu ^3-48 \nu ^2+290 \nu +84\right)\\
    &\quad+\alpha ^2 \left(28 \nu
   ^4-189 \nu ^3+54 \nu ^2+290 \nu -3\right)-\alpha ^5 (\nu +42)+3 \alpha ^6\\
   &\quad+3 \alpha  \left(24 \nu ^4-63 \nu ^3-16 \nu ^2-11 \nu -14\right)+28
   \nu ^4-3 \nu ^3+21 \nu ^2-\nu +3\Big)\,.
\end{align*}
The first two current multipole moments in the $\phi_1$-direction are given by
\begin{align*}
    J^{(1)}&=\frac{\pi R^3}{2}\frac{\sqrt{2} \nu \left(\alpha ^2+\alpha  (\nu -6)+\nu +1\right) \sqrt{\alpha ^2+2 \alpha -\nu ^2+1}}{(\alpha -1)^2 (1+\nu^2)^{3/2} (\alpha
   -\nu +1)^2}\,,\\
   S^{(1)}_3&=\frac{2J^{(1)}}{\pi}\frac{R^2}{2 (\alpha -1)^2 (1+\nu^2) (\alpha -\nu +1) \left(\alpha ^2+\alpha
    (\nu -6)+\nu +1\right)} \Big(\alpha ^3 \left(3 \nu ^2+14\right)\\
    &\quad +\alpha ^2 \left(3 \nu ^3-35 \nu ^2+62 \nu +14\right) +\alpha ^4 (\nu -15)+\alpha ^5+5
   \alpha  \left(2 \nu ^3-7 \nu ^2-3\right)+3 \nu ^3+3 \nu ^2+\nu +1\Big)\,,
\end{align*}
and those in the $\phi_2$-direction are given by
\begin{align*}
    J^{(2)}&=\frac{\pi R^3}{2}\frac{2 \sqrt{2} \sqrt{\alpha } \nu  \sqrt{\alpha ^2+2 \alpha -\nu ^2+1}}{(\alpha -1)^2 (1+\nu)^{3/2} (\alpha -\nu +1)}\,,\\
    S_3^{(2)}&=\frac{2J^{(2)}}{\pi}\frac{3 R^2 \left(-\alpha ^2 (\nu +1)+\alpha ^3+\alpha  \left(2 \nu ^2-6 \nu -1\right)+2 \nu ^2-\nu +1\right)}{(\alpha -1)^2 (1+\nu^2) (\alpha
   -\nu +1)}\,.
\end{align*}
These multipole moments coincide with the multipole moments of the one-angular momentum black ring when $\alpha=0$.

\subsection{Multicentered geometries}\label{sec:BWmultipoles}

As a more elaborate example of calculating mass and current multipoles using the ACMC formalism, we turn to the general class of five-dimensional supersymmetric multicentered geometries \cite{Bena:2007kg}. In particular, we will focus on smooth, horizonless geometries --- a subfamily of the most general multicentered geometries. These smooth geometries are solutions to five-dimensional supergravity; depending on the moduli, they can have either $\mathbb{R}^{4,1}$ or  $\mathbb{R}^{3,1} \times S^1$ asymptotics. The four-dimensional multipole moments of such smooth multicentered geometries with $\mathbb{R}^{3,1} \times S^1$ asymptotics were calculated in \cite{Bena:2020see,Bena:2020uup,Bianchi:2020bxa,Bianchi:2020miz}; here, we will calculate their five-dimensional mass and current multipoles when they have $\mathbb{R}^{4,1}$ asymptotics. With such asymptotics, they can be considered to be microstates corresponding to the supersymmetric BMPV black hole.

\subsubsection{The metric}

The five-dimensional metric is given by \cite{Bena:2007kg}:
\be \label{eq:BWmetric} ds^2 = -(Z_1 Z_2 Z_3)^{-2/3}(dt+k)^2 + (Z_1 Z_2 Z_3)^{1/3}\left( V^{-1}(d\psi + A)^2 + V  ds_3^2\right),\ee
where $ds_3^2 = d\rho^2 + \rho^2d\ttheta^2 + \rho^2\sin^2\ttheta d\phi^2$ is the usual flat metric on the $\mathbb{R}^3$ base space. The metric is entirely defined by specifying eight harmonic functions $(V, K^I, L_I,M)$, with $I=1,2,3$, on this $\mathbb{R}^3$, which are related to the metric functions as:
\begin{align}
\label{eq:ZIdef} Z_I & = L_I + \frac12 C_{IJK} \frac{K^J K^K}{V},\\
\mu &= M + \frac{1}{2V} K^I L_I + \frac{1}{6V^2}C_{IJK}K^I K^J K^K,\\
k & =\mu(d\psi +A) + \omega ,\\
\vec{\nabla}\times \vec{\omega} &= V \vec{\nabla} M  - M \vec{\nabla} V + \frac12\left( K^I\vec{\nabla} L_I - L_I \vec{\nabla} K^I\right),\\ 
\label{eq:AV} \vec{\nabla}\times \vec{A} &= \vec{\nabla} V,
\end{align}
where $C_{IJK}$ characterizes the five-dimensional supergravity theory in which we work; we will take $C_{IJK} = |\epsilon_{IJK}|$ (the STU model). Note that the last two equations determining $\omega$ and $A$ are equations on the (flat) $\mathbb{R}^3$ base.

The eight harmonic functions $H=(V, K^I, L_I,M)$ are determined by the locations of their poles in $\mathbb{R}^3$, $\vec{\rho}_i$ ($i=1,\cdots, N$), which are commonly known as ``centers''. The coefficients $h^i$ are also called the charges associated to the center $i$, collectively denoted in the charge vector $\Gamma^i$:
\be \Gamma^i = \left(v^i, k_1^i, k_2^i, k_3^i, l_1^i, l_2^i, l_3^i, m^i\right),\ee
and the constant terms in the harmonic functions are the moduli at infinity $h$:
\be h = \left( v^0, k_1^0, k_2^0, k_3^0, l_1^0, l_2^0, l_3^0, m^0 \right),\ee
so that the harmonic functions are given in totality by:
\be H = h + \sum_{i=1}^N \frac{\Gamma^i}{\rho_i},\ee
where $\rho_i\equiv |\vec{\rho}-\vec{\rho}_i|$ is the distance in $\mathbb{R}^3$ to the $i$'th center.

We will consider only configurations where the centers are on the $z$-axis (so that the metric admits biaxial symmetry). We can then order the centers such that $z_i>z_j$ when $i<j$, and then an explicit expression for $\omega$ is given by \cite{Bena:2007kg}:
\be \omega = \sum_{i<j} \omega_{ij} d\phi,\ee
with each pair of centers $(i,j)$ contributing:
\begin{align}
\nn \omega_{ij} &=  \frac{\langle \Gamma^i,\Gamma^j\rangle}{|z_i-z_j|}\left( \frac{z_i -\rho\cos\ttheta}{\sqrt{\rho^2 + z_i^2 - 2 \rho z_i \cos\ttheta}} - (z_i\leftrightarrow z_j) \right.\\
\label{eq:omegaij} & \left.+ \frac{\rho^2 + z_iz_j -\rho(z_i+z_j)\cos\ttheta}{\sqrt{\rho^2 + z_i^2 - 2 \rho z_i \cos\ttheta}\sqrt{\rho^2 + z_j^2 - 2 \rho z_j \cos\ttheta}} -1  \right) ,\end{align}
where we have used the symplectic product of two charge vectors:
\be \langle \Gamma^i, \Gamma^j \rangle \equiv m^i v^j - \frac12 k_I^i l_I^j - (i\leftrightarrow j).\ee

To avoid closed time-like curves at any center $i$, necessary conditions that the positions of the centers and their charges must satisfy are
the so-called bubble equations:
\be \label{eq:bubbleeqs} \sum_{j\neq i} \frac{\langle\Gamma^i, \Gamma^j\rangle}{ |\vec{\rho}_i - \vec{\rho}_j|} = \langle h, \Gamma^i\rangle\, .\ee
If we further demand that the five-dimensional solution is completely \emph{smooth}, then at each center the $L_I$ and $M$ charges are determined in terms of the $K^I, V$ charges as \cite{Bena:2007kg}:\footnote{In particular, this condition makes the centers smooth Gibbons-Hawking centers; there are other possibilities where the centers are smooth but not G-H centers, e.g. supertube centers.}
\be \label{eq:smoothcenter} l_I^i = -\frac12 \frac{ C_{IJK}k_J^i k_K^i}{v^i}, \qquad m^i = \frac1{12} \frac{C_{IJK}k^i_I k^i_J k^i_K}{v_i^2}.\ee

To achieve five-dimensional $\mathbb{R}^{4,1}$ asymptotics, it is necessary to demand $v^0=0$ and $\sum_i v^i = 1$. For simplicity, we will always take the moduli to be:
\be \label{eq:modulichoice} h = \left( v^0, k_1^0, k_2^0, k_3^0, l_1^0, l_2^0, l_3^0, m^0 \right) = (0, 0,0,0, 1,1,1, m^0),\ee
where $m^0$ is determined by solving the sum of the bubble equations (\ref{eq:bubbleeqs}), or equivalently by demanding that $\mu\rightarrow 0$ as $\rho\rightarrow \infty$:
\be \label{eq:m0eq} m^0 = -\frac12 \sum_{I=1}^3 \sum_i k^i_I.\ee
The choice (\ref{eq:bubbleeqs}) and (\ref{eq:m0eq})  will always be supplemented by the condition
\be \label{eq:sumvi} \sum_i v^i = 1.\ee

Finally, it is useful to mention that the system above has a gauge freedom where the harmonic functions are shifted as \cite{Bena:2007kg}:
\be \label{eq:gaugetransf} \begin{aligned} K^I &\rightarrow K^I + c^I V,\\
L_I &\rightarrow L_I - C_{IJK} c^J K^K -\frac12 C_{IJK}c^Jc^K V,\\
M &\rightarrow M - \frac12 c^I L_I  + \frac{1}{12}C_{IJK}\left(c^I c^Jc^K V + 3c^I c^J K^K\right),\end{aligned}\ee
for arbitrary constants $c^I$.
Physical quantities (such as charges or multipole moments) should then always be gauge-invariant under the transformation (\ref{eq:gaugetransf}).

\subsubsection{Connection with ACMC coordinates}
Asymptotically, for $\rho\rightarrow \infty$, we have (using (\ref{eq:modulichoice}), (\ref{eq:m0eq}), and (\ref{eq:sumvi})):
\be Z_I\rightarrow 1, \qquad V \rightarrow \frac{1}{\rho}, \qquad A \rightarrow \cos\ttheta d\phi,\qquad (\mu,\omega)\rightarrow 0.\ee
With the coordinate change $\rho = r^2/4$, the metric asymptotes to:
\be ds^2 \rightarrow -dt^2 + dr^2 +  \frac{r^2}{4}\left(d\ttheta^2 + \sin^2\ttheta d\phi^2 + (d\psi + \cos\ttheta d\phi)^2\right).\ee
Finally, we need the angle redefinition:
\be \label{eq:angleredef} \ttheta = 2\theta, \qquad \psi = \phi_1 + \phi_2, \qquad \phi = \phi_2-\phi_1,\ee
to reach the standard five-dimensional biaxial coordinates $(r,\theta,\phi_1,\phi_2)$ used in Section \ref{sec:biaxial}.

In four dimensions, these multicentered geometries are automatically four-dimensional ACMC-$\infty$ in the $(\rho,\ttheta,\phi)$ coordinates \cite{Bena:2020uup}. The situation in five dimensions is similar: after the transformations $\rho = r^2/4$ and (\ref{eq:angleredef}), the five-dimensional metric expansion is automatically ACMC-$\infty$; see Section \ref{sec:MGcurrentACMC}.

\subsubsection{Mass multipoles}
To determine the mass multipoles, we need to find a general asymptotic expansion of the $Z_I$ functions, using similar techniques as described in detail in \cite{Bena:2020uup,Bah:2021jno}. The main difference with  \cite{Bena:2020uup,Bah:2021jno} is that both $K^I\rightarrow 0$ and $V\rightarrow 0$ as $r\rightarrow \infty$, which means the $K^J K^K/V$ term contributes in a non-trivial way in this expansion.

We use (\ref{eq:ZIdef}), the choice of moduli (\ref{eq:modulichoice}), and the condition (\ref{eq:smoothcenter}) for smooth centers. Then, introducing the gauge-invariant fluxes between pairs of centers $\Pi^I_{ij}$ defined as \cite{Bena:2007kg}:
\be \Pi^I_{ij} = \frac{k^j_I}{v^j} - \frac{k^i_I}{v^i},\ee
we obtain an expression for $Z_I$ as:
\be \label{eq:ZIexpansion} Z_I = 1 - \sum_{\ell =0}^\infty \frac{1}{\rho^{\ell+1}}\left[\left( \frac12\sum_{i<j} v^i v^j C_{IJK}\Pi^J_{ij} \Pi^K_{ij}\,\mathcal{P}^{(\ell)}_{ij}(z_k, v^k)\right)P_\ell(\cos\ttheta) + \text{(subleading angular dependence)}\right],\ee 
where $\mathcal{P}^{(\ell)}_{ij}(z_k, v^k)$ is polynomial in (all) the $z_k$ and $v^k$'s, and in particular is homogeneous in the $z_k$'s of degree $\ell$ \emph{and} homogeneous in the $v^k$'s of degree $\ell$. (Additionally, by construction $\mathcal{P}^{(\ell)}_{ij}=\mathcal{P}^{(\ell)}_{ji}$.) For example:
\begin{align} \mathcal{P}^{(\ell=0)}_{ij}  &= 1,\\
\mathcal{P}^{(\ell=1)}_{ij}  &= z_i(1-v^i) + z_j(1-v^j) - \sum_{k\neq i,j} v^k z_k = z_i\sum_{k\neq i} v^k + z_j\sum_{k\neq j} v^k - \sum_{k\neq i,j} v^k z_k.
\end{align}
However, it is not clear what the closed-form expression is for $\mathcal{P}^{(\ell)}_{ij}$ for general $\ell>1$.

The coordinate transformations discussed in the previous section transforms:
\be \frac{P_{\ell}(\cos\ttheta)}{\rho^{\ell+1}} = \frac{4^{\ell+1} P_{\ell}(\cos2\theta)}{r^{2\ell+2}},\ee
from which it then follows that the mass multipoles $M_{\ell}$, for even $\ell$, are given by:
\be \begin{aligned} M_{2n}& = \frac{3\pi}{8} (-1)^{n} 4^{n+1} \left(\frac{1}{3}\right) \sum_I \left(- \sum_{i<j} v^i v^j C_{IJK}\Pi^J_{ij} \Pi^K_{ij}\,\mathcal{P}^{(n)}_{ij}(z_k, v^k)\right)\\
&= \pi (-1)^{n+1}2^{2n-1}  \sum_I \sum_{i<j} v^i v^j C_{IJK}\Pi^J_{ij} \Pi^K_{ij}\,\mathcal{P}^{(n)}_{ij}(z_k, v^k),
\end{aligned}\ee
(and $M_{2n+1} = 0$ as always in a biaxially symmetric geometry).
We did not explicitly mention the three gauge fields $A^I$ (as well as two scalars) that are also present in this solution; the three electric charges associated to these gauge fields are \cite{Bena:2007kg}:\footnote{We are using units where $G=1$. By contrast, \cite{Bena:2007kg} uses units where $G=\pi/4$, for which $M = Q_1+Q_2+Q_3$. Note that in both cases, the normalization for the charges is such that $Z_I\sim 1 + Q_I/r^2+\cdots$.}
\be Q_I = - 2C_{IJK} \sum_{i<j} v^i v^j C_{IJK}\Pi^J_{ij} \Pi^K_{ij},\ee
so that:
\be M_0 = M = \frac{\pi}{4}\left(Q_1 + Q_2 + Q_3\right),\ee
as appropriate for a supersymmetric solution.

\subsubsection{Current multipoles and the ACMC conditions on other metric components}\label{sec:MGcurrentACMC}

The current multipoles are determined by the asymptotic behaviour of $\mu$ and $\omega$ in the rotation form $k$; note that since $Z_I\rightarrow 1$ and $\mu,\omega\sim 1/r$, the $Z_I$ factors will not contribute to the current multipoles. As a result, the $g_{0\psi}$ component of the metric has an expansion of the form:
\be\label{eq:muexp} g_{0\psi} = \mu =  \sum_{n=0}^\infty \frac{(-1)^{n+1}}{\pi 4^n \rho^{n+1}}\left(\left(  \left[S_{2n+1}^{(1)}+S_{2n+1}^{(2)}\right] + \left[S_{2n+1}^{(2)}-S_{2n+1}^{(1)}\right] \cos\ttheta\right) P_{n}(\cos\ttheta)+\cdots \right) ,\ee
where we used the relation (\ref{eq:angleredef}) between the ACMC-$\infty$ angles $\phi_{1,2}$, and the $\cdots$ are subleading angular dependences. In principle, from this expression, the linear combination of multipoles $S_{2n+1}^{(2)}-S_{2n+1}^{(1)}$ can be read off at each order $n$.\footnote{The other linear combination of current multipole moments cannot be read off unambiguously only from $g_{t\psi}$, since the subleading angular dependences to the $S_{\ell}^{(2)}-S_{\ell}^{(1)}$ term will mix with the angular dependence of the $S^{(1)}_{\ell}+S^{(2)}_{\ell}$ term. To extract both combinations of current multipole moments, also the $g_{0\phi}$ term is needed.}
We have checked that the expression for the angular momentum combination $S^{(2)}_{1}-S^{(1)}_{1}$ agrees with \cite{Bena:2007kg}; however,
we leave the explicit extraction of the rest of the current multipoles to future work.

\bigskip

Since the harmonic function $V$ has an expansion given by:
\be V = \sum_{\ell=0}(\sum_i v^iz_i^\ell) \frac{P_{\ell}(\cos\ttheta)}{r^{\ell+1}},\ee
it follows from (\ref{eq:AV}) that $A = A_{\phi}\, d\phi$ is given by:
\be\label{eq:Aexpansion} A_{\phi} = \cos\ttheta + \sum_{\ell=1}  \frac{(\sum_i v^iz_i^\ell)}{r^\ell}\left[\cos\ttheta P_\ell(\cos\ttheta) - P_{\ell-1}(\cos\ttheta)\right] 
.\ee
A similar expression for $\omega=\omega_\phi \, d\phi$ was derived in \cite{Bena:2020uup}:
\be\label{eq:omegaexpansion} \omega_\phi = \sum_{i<j} \omega_{ij} =  \sum_{\ell=1} \sum_{i<j} \frac{\langle \Gamma^i, \Gamma^j\rangle }{|z_i-z_j|} \frac{z_j^\ell-z_i^\ell}{r^\ell}\left[\cos\ttheta P_\ell(\cos\ttheta) - P_{\ell-1}(\cos\ttheta) + \cdots\right],
\ee
where the $\cdots$ denote subleading angular dependences.

From the metric function expansions (\ref{eq:ZIexpansion}), (\ref{eq:muexp}), (\ref{eq:Aexpansion}), and (\ref{eq:omegaexpansion}), it can be checked that the metric (after the transformation $\rho=r^2/4$ together with the angle redefinition (\ref{eq:angleredef})) is automatically ACMC at each order, i.e. the metric is ACMC-$\infty$.

\section*{Acknowledgments}
We would like to thank I. Bena, P. Cano, T. Li, and T. Shiromizu for interesting discussions.
We also thank Claudio Gambino, Paolo Pani, and Fabio Riccioni, authors of \cite{Gambino:2024uge}, for bringing to our attention our oversight of the ``stress multipoles'' in our analysis; v2 of this paper indicates the places in which these multipoles would feature.
D.R.M. is supported by Odysseus grant G0H9318N of FWO Vlaanderen.
This work is also partially supported by the KU Leuven C1 grant ZKD1118 C16/16/005.

\appendix

\section{STF and ASTF Tensors}\label{app:STF}

Symmetric and trace free (STF) tensors were used extensively in Thorne's original ACMC formalism and definition of multipoles in four dimensions \cite{Thorne:1980}. Using the notation $A_\l = a_1 a_2\cdots a_\l$, the STF part of a tensor $T_{A_\l} = T_{a_1\cdots a_\l}$ is denoted as $[\T_{A_\l}]^{\text{STF}}$. Such an STF tensor of rank $\l$, i.e. $\T_{A_\l} = [\T_{A_\l}]^{\text{STF}}$, is completely symmetric:
\be [\T_{a_1 a_2\cdots a_\l}]^{\text{STF}} = [\T_{(a_1 a_2\cdots a_\l)}]^{\text{STF}},\ee
and has all vanishing traces, i.e. for all $1\leq i,j\leq \l$:
\be \delta^{a_i a_j}[\T_{a_1 a_2\cdots a_\l}]^{\text{STF}} = 0 .\ee
The STF part of a tensor can be obtained explicitly by symmetrizing and subtracting traces; for example:
\be [\T_{ab}]^{\text{STF}} = T_{(ab)} - \frac1{d}\delta_{ab}T_{dd}; \quad  [\T_{abc}]^{\text{STF}} = \T_{(abc)} -\frac1{d+2}\left( \delta_{ab}\T_{(ddc)} + \delta_{ac}\T_{(dbd)} + \delta_{bc}\T_{(add)}\right),\ee
for tensors living in $d$ spatial dimensions. An STF tensor $\T_{A_\l} = [\T_{A_\l}]^{\text{STF}}$ of rank $\l$ has number of independent components given by:
\begin{equation}
    \#[[\T_{A_\l}]^{\text{STF}}] = \binom{d+\l-1}{\l}-\binom{d+\l-3}{\l-2},
\end{equation}
in $d$ spatial dimensions; in our case $d=4$, so  $\#[[\T_{A_\l}]^{\text{STF}}] = (1+\l)^2$.

STF tensors arise naturally in harmonic solutions, since the most general solution to $\Laplace f = 0$ that vanishes for $r\rightarrow\infty$ is given by (in four spatial dimensions):
\be f = \sum_{\l=0}^\infty \F_{A_\l} \partial_{A_\l} \left( \frac{1}{r^2}\right) = \sum_{\ell=0}^\infty \F_{A_\l} \left( \frac{1}{r^2}\right)_{,\,A_\l} ,\ee
where $\F_{A_\l}$ is STF by construction --- it is completely symmetric as the partial derivatives on $1/r^2$ commute, and it is trace-free as $\partial_{ii}(1/r^2) = 0$.

\medskip

In four spatial dimensions, as opposed to Thorne's three spatial dimensions, we also require a new type of tensor which we call ``anti-symmetric + STF'' (ASTF). Such an ASTF tensor $\D_{bc\, A_\l}$ is defined as being STF on all indices except the first two:
\be [\D_{bc\, a_1\cdots a_\l}]^\text{ASTF} =  [\D_{bc\, (a_1\cdots a_\l)}]^\text{ASTF}, \qquad \delta^{a_ia_j}[\D_{bc\, a_1\cdots a_\l}]^\text{ASTF} = 0,\ee
and antisymmetric on the first two indices:
\be [\D_{bc\, a_1\cdots a_\l}]^\text{ASTF} = [\D_{[bc]\, a_1\cdots a_\l}]^\text{ASTF}.\ee
Additionally, all traces must vanish, so also:
\be \delta^{ba_i} [\D_{bc\, a_1\cdots a_\l}]^\text{ASTF} = \delta^{ca_i}[\D_{bc\, a_1\cdots a_\l}]^\text{ASTF} = 0,\ee
for all $1\leq i\leq \l$; and finally all antisymmetrizations over three indices vanish, so:
\be [\D_{[bc\, a_1] a_2\cdots a_\l}]^\text{ASTF} = [\D_{[bc\, |a_1\cdots a_{i-1}|a_i]a_{i+1}\cdots a_\l}]^\text{ASTF}= [\D_{[bc\,| a_1\cdots |a_\l]}]^\text{ASTF} = 0.\ee
Such ASTF tensors are not necessary in Thorne's three spatial dimensions as then the antisymmetric pair of indices can be exchanged for a single index, $\D_{bc\, A_{\l}} = \epsilon_{bc a_0} \D'_{a_0 A_{\l}}$, thus obtaining a rank $\l+1$ STF tensor $\D'_{ A_{\l+1}}$. By contrast, as we show in this paper, the current multipoles in five dimensions (four spatial dimensions) are naturally described by precisely such an ASTF tensor.

It is tricky to count the number of independent components for a general ASTF tensor;\footnote{Indeed, our counting in \eqref{eq:countingD} of the degrees of freedom for $\D_{bca_1a_2}$ was mistakenly tabulated as 40 in v1 of this paper; we thank the authors of \cite{Gambino:2024uge} for pointing this out.} for the first few ASTF tensors in four spatial dimensions, we have:
\be \label{eq:countingD}
\#[ [\D_{bc}]^\text{ASTF}] = 6 , \qquad \#[ [\D_{bc \,a_1}]^\text{ASTF}]= 16 ,\qquad
\#[ [\D_{bc\,a_1a_2}]^\text{ASTF}] =30,\ee
ASTF tensors arise naturally in parametrizing  solutions to the vector Laplace equation in four spatial dimensions, $\Laplace V_j = 0$; the most general solution that vanishes for $r\rightarrow \infty$ is given by (see (\ref{eq:mostgenerallaplacesolvec})):
\be \label{eq:applaplacevec}
 V_j=\sum_{\l=0}^\infty\B_{A_L}\left(\frac{1}{r^2}\right)_{,\,jA_\l}+\sum_{\l=1}^\infty\C_{jA_{\l-1}}\left(\frac{1}{r^2}\right)_{,\,A_{\l-1}}+\sum_{\l=2}^\infty \varepsilon_{jbca}\D_{bcA_{\l-2}}\left(\frac{1}{r^2}\right)_{,\,a A_{\l-2}},
 \ee
 with all $\B_{A_L}, \C_{A_L}$ STF tensors and all $\D_{bcA_{\l-2}}$ ASTF tensors. It can be verified that the splitting in STF and ASTF parts as in \eqref{eq:applaplacevec} correctly parametrizes the number of degrees of freedom of a general vector $V_j$.\footnote{The most general symmetric tensor $\gamma_{ij}$ solving $\Laplace \gamma_{ij}=0$ decomposes into STF and ASTF parts, but in addition also contains a ``Riemann-symmetric'' (RSTF) part; see \cite{Gambino:2024uge}.}

\section{Geroch-Hansen Formalism in Five Dimensions}\label{app:GH}
The Geroch-Hansen formalism, as generalized in \cite{Tanabe:2010} to five dimensional vacuum spacetimes, assumes asymptotic flatness and the existence of a timelike Killing vector $\xi$ with norm $\lambda = \xi^2$. One can then define the spacelike four-dimensional metric $\hat{h}_{ab}$ as
\be \hat{h}_{ab} = g_{ab} -\lambda^{-1} \xi_a \xi_b .\ee
In coordinates where $\xi = \partial_t$, this three-dimensional spatial manifold is simply any $t=const.$ hypersurface. We then consider a conformal transformation $h_{ab} = \Omega^2 \hat{h}_{ab}$ such that infinity is brought to a single point $\Lambda$. This implies we must have $\Omega\sim 1/r^2$; more precisely, we demand at $\Lambda$ that:
\be \Omega|_{\Lambda} = (D_a \Omega)_\Lambda = 0, \qquad (D_a D_b \Omega)_\Lambda = 2h_{ab},\ee
where $D_a$ is the covariant derivative with respect to the transformed metric $h_{ab}$.

Up to this point, the discussion would have been completely the same for four or five spacetime dimensions. Now, in five spacetime dimensions, \cite{Tanabe:2010} introduce the STF tensors $P_{A_\l}$ recursively as:
\be \label{eq:PsGH} \begin{aligned}
    P &= \Omega^{-1} (1-\sqrt{\lambda}),\\
    P_a &= D_a P,\\
    P_{a_1\cdots a_\l} &= \left[ D_{a_1} P_{a_2\cdots a_\l}  - \frac{(\l-1)^2}{2} R_{a_1a_2}P_{a_3\cdots a_\l}\right]^{\text{STF}}.
\end{aligned}\ee
The mass multipole tensors $\cal{M}_{A_\l}$ in this formalism are then given by:
\be \mathcal{M}_{A_\l} = P_{A_\l}|_{\Lambda}.\ee
The normalization is such that the mass is given by $M = (3\pi/4)   P$ \cite{Tanabe:2010}. Note that the main difference between \eqref{eq:PsGH} in four spatial dimensions compared to the original Geroch-Hansen formalisms is the definition of $P$ in terms of $\lambda$ only --- in Hansen's definition of the monopole tensor, the twist scalar is also present \cite{Hansen:1974}; moreover, the conformal weight of $P$ is $\Omega^{-1}$ (which in three spatial dimensions is $\Omega^{-1/2}$), and the factor multiplying the Ricci tensor term is $(\l-1)^2/2$ (in three spatial dimensions this is $(\l-1)(2\l-3)/2$).

The current multipole tensors are defined by \cite{Tanabe:2010} using the twist two-form $\omega_{ab}$ (in three spatial dimensions, this is a one-form):
\be \omega_{ab} = \hat{\epsilon}_{abcde} \xi^c \hat{\nabla}^d \xi^e,\ee
where the hatted quantities indicate they are taken in the full (untransformed) five-dimensional metric. The vacuum Einstein equations imply $\hat{\nabla}_{[a}\omega_{bc]} =0$, so a twist potential one-form can be defined through $\omega_{ab} = \hat{\nabla}_{[a}\omega_{b]}$. At this point, \cite{Tanabe:2010} specifies biaxial symmetry with (commuting) Killing vectors $l^a, m^a$, and then defines two scalar potentials as:
\be J^{\phi_1} = \frac{\omega_a l^a}{(l_a l^a)^{1/2}}, \qquad J^{\phi_2} = \frac{\omega_a m^a}{(m_a m^a)^{1/2}}.\ee
These then allow two towers of current multipole towers to be defined recursively:
\be J^{\phi_{1,2}}_{a_1\cdots a_\l} = \left[ D_{a_1} J^{\phi_{1,2}}_{a_2\cdots a_\l}  - \frac{(s-1)^2}{2} R_{a_1a_2}J^{\phi_{1,2}}_{a_3\cdots a_\l}\right]^{\text{STF}}.\ee 
The multipole moments are then again these tensors evaluated at $\Lambda$.

\bigskip

A claim in \cite{Tanabe:2010} was that up to the quadrupoles (e.g. $\mathcal{M}_{ab}$), this procedure gives an unambiguous definition of multipoles, but for higher multipoles, the procedure is ambiguous. The reasoning in \cite{Tanabe:2010} is that under a change of conformal factor $\Omega\rightarrow \omega\, \Omega$ (with necessarily $\omega|_\Lambda = 1$), to linear order in $\omega-1$, the multipoles\footnote{We only focus on the mass multipoles here, since this is also the focus of the ambiguity claim in \cite{Tanabe:2010}.} transform as:
\be \mathcal{M}_{A_\l} \rightarrow \mathcal{M}_{A_\l} - s^2 \left[ D_{a_1}\omega \mathcal{M}_{a_2\cdots a_\l}\right]^{\text{STF}},\ee
which is indeed the correct transformation of the multipole tensors under a change of conformal factor --- this simply represents the (correct) transformation of multipoles under a change of spatial origin \cite{Geroch:1970:2}. However, \cite{Tanabe:2010} claim that the \emph{non-linear} part of the transformation of $\mathcal{M}_{A_\l}$ renders the multipoles ambiguous, in particular because this part of the transformation depends on higher-order derivatives of $\omega$ such as $D_aD_b\omega$.

However, this seems not to be true. For example, the transformation of $\mathcal{M}_{abc}$ under the conformal transformation is given by:
\be \mathcal{M}_{abc} \rightarrow  \mathcal{M}_{abc}  - 9  \left[ D_a\omega \,\mathcal{M}_{bc}\right]^{\text{STF}}
 + 18\left[ D_a\omega\, D_b\omega\, \mathcal{M}_c \right]^{\text{STF}} - 6\left[ D_a\omega\, D_b\omega\, D_c\omega \mathcal{M}\right]^{\text{STF}}.
\ee
Note that this is the \emph{full} transformation of $\mathcal{M}_{abc}$, i.e. to all orders in $\omega$.
Clearly, there are no terms proportional to $D_a D_b\omega$ in this transformation, contrary to what was reported in \cite{Tanabe:2010}. Rather, to all orders in $\omega$, the transformation only depends on powers of $D_a\omega$, as it should (and completely analogously to the original Geroch-Hansen case in three dimensions).

In other words, the obstruction to or ambiguity in defining the multipole moments in this Geroch-Hansen fashion, as reported by \cite{Tanabe:2010}, is simply not present. However, we also note that this does not immediately imply that the Geroch-Hansen method is rigorously well-defined in five dimensions. The well-definedness of the original Geroch-Hansen procedure depends crucially on the conformally invariant Laplacian equation of motion that the multipole scalars (i.e. the four-dimensional analogue of $P$, $J^{\phi_i}$) satisfy, which are implied by the vacuum Einstein equations. The appropriate conformally-invariant equation of motion has not yet been shown for $P$, $J^{\phi_i}$. Moreover, in four dimensions, the existence of a conformal factor $\Omega$ and compactified metric $h_{ab}$ that are smooth at $\Lambda$ for arbitrary asymptotically flat four-dimensional vacuum spacetimes (a result by Beig and Simon \cite{Beig:1980,Beig:1981,Beig:1983,Beig:1981:2} and independently Kundu \cite{kundu1981multipole,kundu1981analyticity}) is key for the conformal compactification lying at the heart of the Geroch-Hansen formalism. Although it is reasonable to assume the generalization of these results to five dimensions is possible, to the best of our knowledge they have not yet been shown explicitly. Similar mathematical properties of four-dimensional vacuum spacetime metrics were used in G\"ursel's proof \cite{Gursel:1983} of the equivalence between the Geroch-Hansen and Thorne definitions of multipoles (up to an unimportant normalization constant), so it is also not immediately clear how such an equivalence would generalize to five dimensions.

\bigskip

In our paper, we focus on generalizing Thorne's ACMC formalism for mass and current multipoles to five dimensions. In addition to typically being computationally more practical than the Geroch-Hansen formalism, Thorne's arguments readily generalize to five dimensions (as we show) without the need for assuming any additional (smoothness) properties of the metric or its conformal compactification. It would certainly be an interesting future direction to rigorously complete the program of \cite{Tanabe:2010} of generalizing also the Geroch-Hansen formalism for mass and current multipoles to five dimensions.\footnote{Note that the program of \cite{Tanabe:2010} should also be generalizable to spacetimes without the biaxial symmetry given by $l^a,m^a$.}

\bibliographystyle{toine}
\bibliography{multipoles}

\end{document}